\documentclass{aa}  

\usepackage{txfonts}
\usepackage{booktabs}
\usepackage[version=4]{mhchem}
\usepackage{graphicx}
\usepackage{color}
\usepackage{hyperref}

\newcounter{chem}
\newcounter{temp}
\newenvironment{chequation}{%
  \setcounter{temp}{\value{equation}}%
  \setcounter{equation}{\value{chem}}%
}{%
  \setcounter{chem}{\value{equation}}%
  \setcounter{equation}{\value{temp}}%
}
\newcommand{\rev}[1]{#1}
\begin{document}

   \title{Hydrogenation of acetaldehyde on interstellar ice analogs reveals limited destruction}

   \author{G. Molpeceres
          \inst{1}
          \and
          T. Nguyen
          \inst{2}
          \and
          Y. Oba
          \inst{2}
          \and
          N. Watanabe
          \inst{2}
          }

   \institute{Departamento de Astrofísica Molecular, Instituto de Física Fundamental (IFF-CSIC), C/ Serrano 121, 28006 Madrid, Spain\\
              \email{german.molpeceres@iff.csic.es}
         \and
         Institute of Low Temperature Science, Hokkaido University N19W8, Kita-ku, Sapporo, Hokkaido 060-0819 Japan\\
             \email{nguyenthanh@lowtem.hokudai.ac.jp}
             }

   \date{\today; \today}

 
  \abstract
   {Acetaldehyde \ce{CH3CHO} is one of the most abundant interstellar complex organic molecules and its hydrogenation has important implications in several fundamental processes of interstellar chemistry, like deuterium fractionation, reactive desorption or the relation between organic functional groups of detected molecules. }
   {We sought to determine which are the main hydrogenation paths of \ce{CH3CHO}. As a partially unsaturated molecule, \ce{CH3CHO} can have links with more hydrogenated species, like ethanol (\ce{C2H5OH}) or with more unsaturated ones, like ketene (\ce{H2CCO}). }
   {We used highly accurate quantum chemical calculations to determine the reaction rate constants for the \ce{CH3CHO + H/D reaction}. We later study, using more approximated methods, the fate of the majoritarian product of the reaction, the acetyl radical \ce{CH3CO} after subsequent reaction with hydrogen or deuterium atoms. Our theoretical results are confronted against our experiments on the hydrogenation and deuteration of \ce{CH3CHO} ice.}
   {We find that acetaldehyde resists hydrogenation, with only a 10\% of conversion to products different than \ce{CH3CHO}. This is due to a predominance of H-abstraction at the HCO moiety, with reaction rate constants up to four orders of magnitude higher than the next possible reaction channel, that is hydrogenation at the aldehydic carbon. The formed \ce{CH3CO} radical experiences barrierless or nearly barrierless reactions in all possible reaction positions, reforming \ce{CH3CHO} and creating a closed loop that protects the molecule against hydrogenation. We constrain the branching ratios for the second reaction from experiments. Our experiments agree with the calculations and from the combination of both we can explain the presence of \ce{H2CCO}, CO, CH4, \ce{C2H5OH}, \ce{H2CO} or \ce{CH3OH} as minor products at the end of the reaction. We provide recommendations for future modeling efforts.}
   {Our results show limited destruction of acetaldehyde, reinforcing the vision of this molecule as an abundant and resilient COM. From the experiments, we are not able to observe the reactive desorption of this molecule. Our results align with other modeling works, showing that the link between \ce{CH3CHO} and \ce{C2H5OH} is not direct. Finally, our results can explain the excess of \ce{CH3CDO} found in prestellar cores.}


   \maketitle
%

\section{Introduction} \label{sec:intro}

The emergence of chemical complexity in the interstellar medium (ISM) is one of the most fascinating topics in astrochemistry. Although the concept of ``complexity'' is generally ill-defined, a molecule composed of six or more atoms, with at least one being carbon, is typically considered an interstellar complex organic molecule \citep{Herbst2009}. Given the harsh conditions in the ISM, the formation mechanisms of these molecules are vastly different from those on Earth. In the ISM, particularly within the ice-covered dust grains that account for approximately 1\% of its mass \citep{bohlin_survey_1978}, many molecules are formed through radical chemistry.

Significant effort has been devoted to studying the synthesis of complex organic molecules (COMs) on interstellar ices \citep[][to give a few representative examples]{Watanabe2002,Fuchs2009,Rimola2014,Garrod2013,Simons2020,Jin2020,Garrod2022,enrique-romero_quantum_2022,ferrero_formation_2023, Molpeceres2024com}. Although the processing reactions of COMs, those that convert or destroy already-formed COMs, have also garnered attention, they remain comparatively underexplored. Examples in the literature include the systematic study of COM hydrogenation \citep{Alvarez-Barcia2018}, the hydrogenation of heterocycles \citep{Miksch2021}, and the processing of formic acid and thioformic acid \citep{Molpeceres2022}. However, the chemical processing of COMs on interstellar grains has important implications for the evolution of chemical complexity. Among these, hydrogen abstraction (H-abstraction) reactions play a pivotal role. These reactions involve a radical (e.g., H, OH, \ce{NH2}) abstracting a hydrogen atom from a COM, producing a closed-shell hydrogenated molecule (e.g., \ce{H2}, \ce{H2O}, \ce{NH3}) and leaving behind a parent radical of the COM. For instance, we recently demonstrated that for hydroxylamine (\ce{NH2OH}), H-abstraction reactions, followed by O- or N-addition, can account for the apparent disappearance of this molecule in the ISM \citep{molpeceres_processing_2023}. Besides, following similar arguments, deuterium enrichment of molecules and COMs (See, for example, the review of \citet{Ceccareli2014}), is enhanced through H-abstraction reactions, following a cycle of H-abstraction and D-addition \citep{Molpeceres2022, Nguyen2021b,Lamberts2017b}. Furthermore, the same cycle can fuel the process of reactive desorption to return the ice-synthesized molecules to the gas phase. This is due to the increase of reactive events for a fixed probability of chemical desorption per reaction \citep{garrod_non-thermal_2007}, as it was evinced by chemical desorption experiments by some of us \citep{Oba2018,Nguyen2020} on \ce{H2S} or \ce{PH3}. Clearly, H-abstraction reactions are critical components of modern astrochemical reaction networks and warrant more detailed investigation.

With advancements in astrochemistry, the pool of molecules requiring detailed study is growing not only in number but also in complexity. For instance, considering the examples of \ce{H2S} and \ce{PH3} mentioned above, these molecules have relatively simple reaction networks \citep{Lamberts2017b,Molpeceres2021,Nguyen2021b}. Starting from their atomic precursors (S and P), the fully hydrogenated molecule is the most stable product. Additionally, H-abstraction reactions exhibit increasing activation energies, while H-addition remains barrierless. These traits make such molecules easier to detect in experimental setups, as the reaction products are limited to one stable product and non-detectable reactive radicals. However, the complexity increases significantly when a carbon atom is introduced, as in carbon monosulfide (CS). Adding carbon leads to reaction networks with approximately 15 reactions en route to forming \ce{CH3SH} \citep{lamberts_interstellar_2018, Nguyen2023}, including irreversible pathways such as \ce{CH3 + H2S}. This highlights the intricate nature of hydrogenation sequences. Studying these networks for complex organic molecules (COMs) routinely identified in cold environments \citep{bacmann_detection_2012,Bacmann2014,Jimenez-Serra2021,JImenezSerra2016,megias_complex_2022,Scibelli2021} becomes increasingly challenging, whether approached experimentally or theoretically.

The molecule examined in this paper, acetaldehyde (\ce{CH3CHO}), is one of the most abundant and extensively studied complex organic molecules (COMs) that can form on ice grains \citep{Jin2020,Lamberts2019,ferrero_formation_2023,Molpeceres2024com}, albeit with some challenges \citep{enrique-romero_theoretical_2021, Perrero2023}. Acetaldehyde can also be synthesized in the gas phase \citep{vazart_gas-phase_2020}. Although \ce{CH3CHO} qualifies as a COM, its reaction with hydrogen remains experimentally tractable \citep{Bisschop2007}, yet interpreting the experimental results is far from straightforward. Hydrogenation of \ce{CH3CHO} reportedly produces a mixture of acetaldehyde (\ce{CH3CHO}), ethanol (\ce{C2H5OH}), methane (\ce{CH4}), carbon monoxide (\ce{CO}), formaldehyde (\ce{H2CO}), and methanol (\ce{CH3OH}). However, the latter three products are linked through a sequential hydrogenation pathway \citep{Watanabe2002}. Meanwhile, \ce{CH4} and \ce{C2H5OH} could, in principle, arise from alternative pathways rather than the direct hydrogenation of \ce{CH3CHO}. This complexity underscores the need to better understand the precise mechanism of \ce{CH3CHO} hydrogenation.\footnote{In this context, ``hydrogenation'' refers broadly to reactions involving H atoms, with distinctions between H-abstraction and H-addition noted where relevant.} Such a detailed understanding is essential for proposing potential deuteration pathways and assessing the likelihood of reactive desorption, as discussed earlier. 

A mechanistic understanding of this reaction can only be achieved through a synergistic combination of accurate theoretical calculations and experimental work. This is the approach we adopted in this study paper where we integrate quantum chemical calculations of the hydrogenation of \ce{CH3CHO}, incorporating a rigorous treatment of nuclear quantum effects, with tailored experiments on the \ce{CH3CHO + H} reaction. Furthermore, our work aims to unify the existing knowledge on this reaction and related reactions, such as \ce{CO + H} and \ce{H2CCO + H} \citep{Watanabe2002, Fuchs2009, mondal_is_2021, Fedoseev2022, ibrahim_formation_2022, ferrero_formation_2023}, by resolving discrepancies where they exist or providing justified support where they align. The structure of this manuscript is as follows: In Section \ref{sec:methods}, we provide a concise overview of our methodology. Section \ref{sec:results} is dedicated to the presentation of our findings. In Section \ref{sec:discussion}, we discuss the results, placing them in the context of experimental, modeling, and broader astrochemical perspectives. Finally, we conclude with key remarks in Section \ref{sec:conclusion}.

\section{Methods} \label{sec:methods}

\subsection{Computational Details}

 Quantum chemical calculations were performed with a twofold objective. First, we use the results of a preliminary exploration to guide our experimental search. Second, when the full experimental procedure was completed, we expanded our quantum chemical calculations to explain the experiments, leading to a self-consistent explanation of \ce{CH3CHO} chemistry under ISM conditions. We can categorize our simulations into three different categories. First, calibration calculations using a small two-water model to test the influence of explicit water molecules in the hydrogenation of \ce{CH3CHO} (Appendix \ref{sec:appA}). Second, derivation of the instanton rate constants \citep{Rommel2011-2,Kastner2014} for H-addition, H-abstraction, and D-addition on \ce{CH3CHO}. Instanton theory is an unification of path integral theory and transition state theory to simulate the quantum nature of the nuclei in the reactant and transition states of a reaction. An instanton represents the most probable tunneling path, that in our formulation is discretized in a number of points or beads, and optimized to a first-order saddle point, like in conventional transition state theory. In this case an optimization of the tunneling path with maximum transition probability (or minimum Euclidean action) is carried out. Third and last, investigation of additional reactions with hydrogen in the acetyl radical (\ce{CH3CO}), the preferred product of the first reaction with H (see Section \ref{sec:results}). Our quantum chemical calculations are designed to prioritize accuracy in the electronic structure calculations over cluster representativity. Acetaldehyde is shown to be a COM with a relatively weak binding with the water surface \citep{molpeceres_desorption_2022, ferrero_acetaldehyde_2022}, which normally leads to a small influence of the water matrix on the reaction rate constants \citep[see, for example][for two recent examples]{ferrero_formation_2023, molpeceres_carbon_2024}. The instanton rate constants were calculated using a sequential cooling scheme starting from a temperature close to the crossover temperature defined as:

 \begin{equation} \label{eq:cooling}
     T_{\textrm{c}} = \frac{\hbar \Omega  }{2\pi k_\text{B}}
 \end{equation}

 \noindent where $\Omega$ corresponds to the imaginary frequency of vibration at the transition state. $T_{\textrm{c}}$ is an estimator of the temperature at which the tunneling effects in the rate constant start to dominate over the purely thermal ones \citep{Gillan1987}. Because the determining quantity to compare with experiments is the reaction rate constant, that we obtain from accurate semiclassical instanton theory calculations, the inclusion of surface effects in our calculations is done \emph{via} the implicit surface approach \citep{Meisner2017}, i.e. making the rotational partition faction equal to one in the calculation of the reaction rate constant. Omitting the ice matrix, allows us to use a significantly better level of theory than when including an ice matrix. In particular, the level of theory used in this work is the combination of the rev-DSD-PBEP86(D4)/jun-cc-pV(T+d)Z level (i.e. double hybrid functional with a large basis set) \citep{Kozuch2011, santra_minimally_2019, Dunning1989, papajak_perspectives_2011, Caldeweyher2019} for the molecular geometries and molecular Hessian and the (U)CCSD(T)/aug-cc-pVTZ \citep{bartlett_manybody_1978, purvis_full_1982,Dunning1989,knowles_coupled_1993, Woon1994} gold standard method for correcting the electronic energies.\footnote{Details on how to run the rev-DSD-PBEP86(D4) calculations with different codes can be found at \url{https://www.compchem.me/revdsd-pbep86-functional}} This level is used for the determination of stationary points in the respective potential energy surface (PES), calculation of the zero point vibrational energies (ZPVE) and the calculation of rate constants in the interaction of H with \ce{CH3CHO}. To incorporate the effect of a polar environment, that is \ce{H2O}, we also computed single point energies using the conductor-like polarizable continuum (CPCM) \Citep{truong_new_1995, barone_quantum_1998, garciarates_effect_2020} model employing a dielectric constant extrapolated for amorphous solid water of $\epsilon$=600 as used in recent astrochemical literature \Citep{nguyen_formation_2019}. We present models with and without CPCM correction. The hydrogenation of the \ce{CH3CO} is investigated with rev-DSD-PBEP86(D4)/jun-cc-pV(T+d)Z, that is, without further coupled cluster refinement, although CPCM single points at the DFT level are also reported. The last set of calculations is carried out using a broken-symmetry DFT formalism, first converging the triplet electronic state and then rotating the orbitals of the incoming reactant (H) to guarantee the correct spin state for the system. The rate constants for the reactions starting with the \ce{CH3CO} as reactant are obtained from classical transition state theory including the tunneling correction from an asymmetric Eckart barrier \citep{Eckart1930}. \rev{The energies coming from pure broken symmetry DFT calculations are subjected to a higher error than when correcting the electronic energies with a high level method.} However, the lack of a reliable single reference solution in the case of biradical species such as the ones partaking in the \ce{CH3CO + H} reaction prompt us to continue with this protocol. The agreement between experiments and theory found in this work shall serve as a proxy benchmark for the validity of our method in this particular case.

 All our electronic structure calculations use the \textsc{ORCA 5.0.4} code \citep{Neese2020}. Geometry optimizations and instanton calculations use a developer version of the \textsc{Dl-Find} code \citep{kae09a} interfaced with \textsc{ChemShell} \citep{Metz2014}.\footnote{Cartesian coordinates supporting the calculations can be found at \url{https://zenodo.org/records/14278652}}

\subsection{Experiments}

All experiments were performed using an experimental apparatus named Apparatus for Surface Reaction in Astrophysics (ASURA). The details of the ASURA system were described in previous studies \citep{WATANABE2006, Nagaoka2007, Nguyen2020, Nguyen2021, Nguyen2023}. In brief, it consists of an ultrahigh vacuum chamber with the basic pressure of 10$^{-8}$ Pa; an aluminum (Al) substrate attached to a He cryostat, an atomic source, a quadrupole mass spectrometer (QMS), and a Fourier transform infrared spectrometer (FTIR). The surface temperature was regulated between 5 and 300~K.

The chemical reactions of solid acetaldehyde (\ce{CH3CHO}) with H (or D) atoms were studied on both an Al substrate and compact amorphous solid water (c-ASW) at low temperatures (typically 10~K). The c-ASW was made by the vapor deposition of water onto the substrate maintained at 110~K, with an estimated thickness of $\sim$ 20 monolayers (ML; 1~ML = 1 $\times$ 10$^{15}$ molecules cm$^{-2}$). The substrate was cooled down to 10~K after the c-ASW layer production. Gaseous \ce{CH3CHO} was pre-deposited onto c-ASW with the deposition rate of 1 ML minute$^{-1}$. The layer thickness of \ce{CH3CHO} was adjusted to 1~ML, which was estimated using the peak area of the CO stretching band at 1728 cm$^{-1}$ and its absorption coefficient of 8.0 $\times$ 10$^{-18}$ cm molecule$^{-1}$ \citep{Bisschop2007}. The pre-deposited \ce{CH3CHO} layer was exposed to the H and D atoms produced via the dissociation of \ce{H2} (or \ce{D2}) in a microwave-discharged plasma in the atomic source chamber. The formed H (D) atoms were cooled to 100~K by multiple collisions with the inner wall of the Al pipe at 100~K \citep{Nagaoka2007}. The flux of H and D atoms was estimated as 5.0 $\times$ 10$^{14}$ and 3.4~$\times$~10$^{14}$~cm$^{-2}$~s$^{-1}$ following the method of \cite{Oba2014}, respectively. During the process of H (or D) exposure on the sample solid, interactions between \ce{CH3CHO} and H (D) atoms were observed in situ using FTIR with a resolution of 2 cm$^{-1}$. The reactants and products desorbed from the surface were monitored by the QMS via the temperature programmed desorption (TPD) method with a ramping rate of 4~K minute$^{-1}$.

\section{Results} \label{sec:results}

\subsection{Theoretical exploration}

\subsubsection{Description of \ce{CH3CHO + H}} \label{sec:ch3cho}

\begin{figure}[h]
    \centering
    \vspace{1.5em}
    \includegraphics[width=0.7\linewidth]{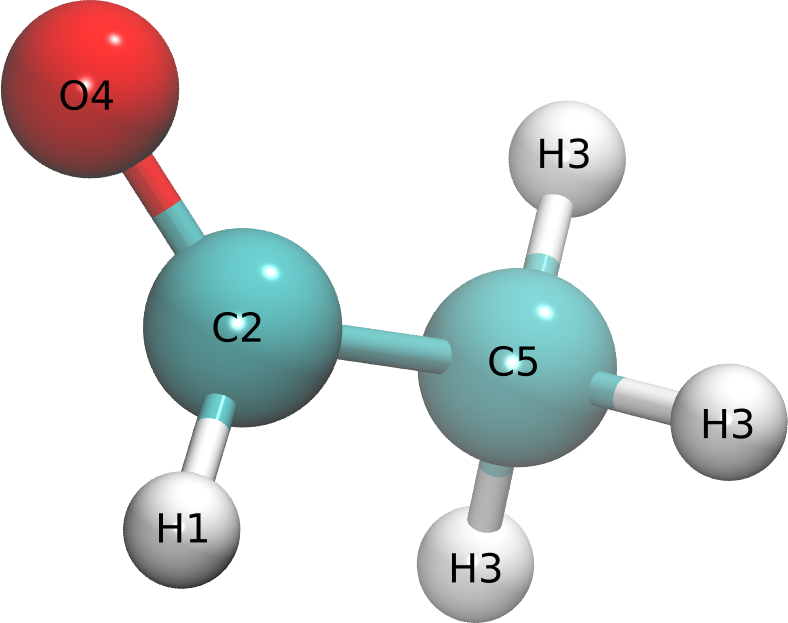} \\
    \vspace{1cm}
    \includegraphics[width=0.5\linewidth]{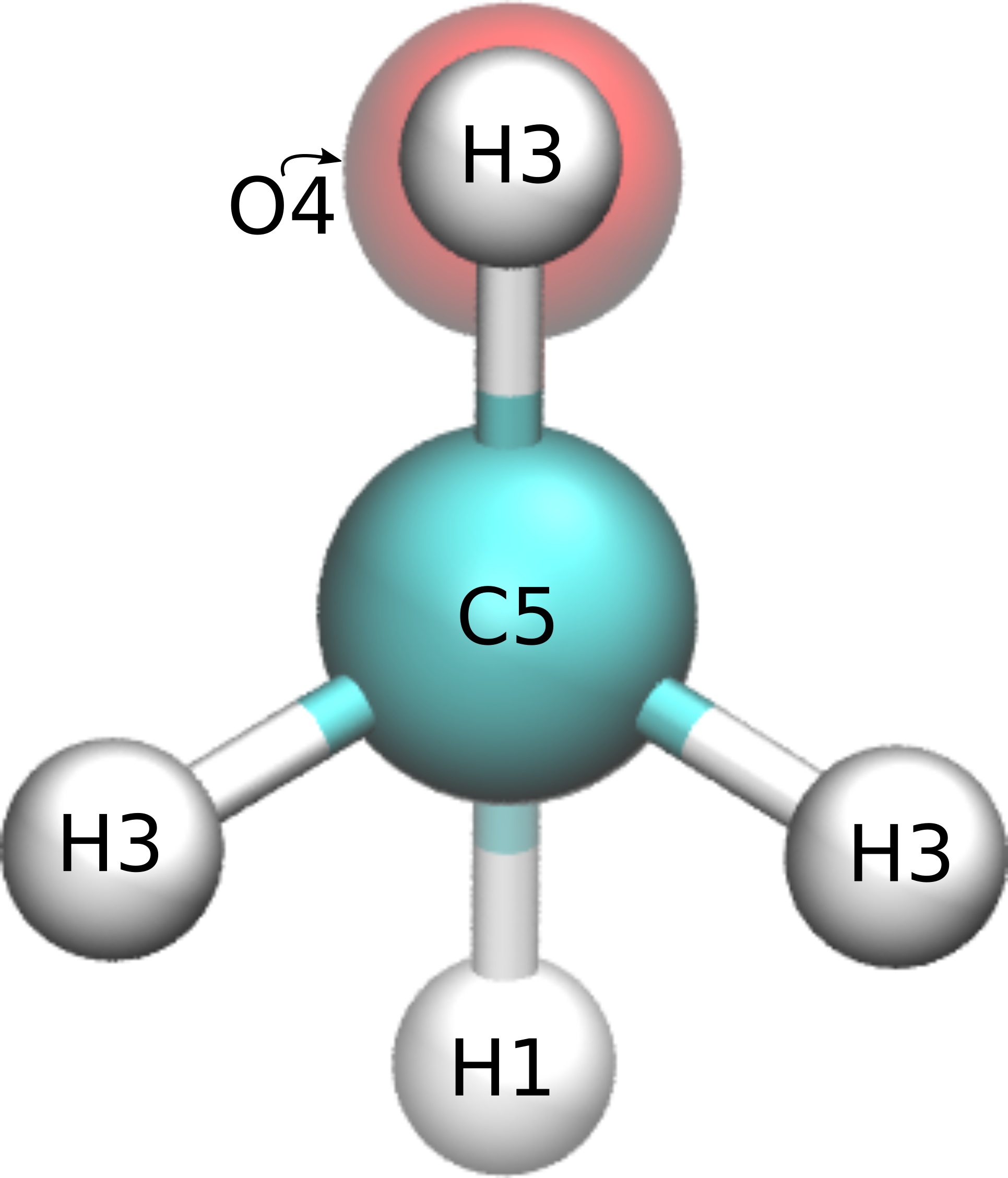} \\
    \caption{Top. Acetaldehyde molecule with the positions where the hydrogenation and deuteration reactions are sampled. Bottom. Newman projection of the molecular model showing the considered \ce{CH3CHO} \textit{syn} conformer}
    \label{fig:theo1}
\end{figure} 

The study of the H-abstraction and addition reactions necessary to rationalize our experiments begins with the evaluation of the energetics for all the possible reactions reaction channels for \ce{CH3CHO + H}. The molecular structure of \textit{syn}-\ce{CH3CHO} and the atomic positions where the reaction can take place are depicted in Figure \ref{fig:theo1}. For each reaction, we assume that energy dissipation occurs via the surface, treating the reactions presented in this study as elementary.The possible reactions are:

\begin{chequation}
\begin{align}
    \ce{CH3CHO + H &-> CH3CO + H2} \label{eq:h1} \\
    \ce{CH3CHO + H &-> CH3CH2O} \label{eq:c2} \\
    \ce{CH3CHO + H &-> CH2CHO + H2} \label{eq:h3} \\
    \ce{CH3CHO + H &-> CH3CHOH} \label{eq:o4} \\
    \ce{CH3CHO + H &-> CH4 + HCO}. \label{eq:c5}
\end{align}
\end{chequation}

\begin{table*}[bt!]
   \begin{center}
    \caption{Reaction enthalpies ($\Delta H^{ \textrm{R}}$, kJ mol$^{-1}$), activation energies ($\Delta H^{ \ddagger}$, kJ mol$^{-1}$), the absolute value of the imaginary frequency of vibration ($\Omega$, cm$^{-1}$) and crossover temperature ($T_{\textrm{c}}$, K) for the reactions of \ce{CH3CHO} with H. We report both the results of the model including implicit interaction with the water matrix (labeled CPCM) and excluding it at the (U)CCSD(T)/aug-cc-pVTZ/revDSD-PBEP86(D4)/jun-cc-pV(T+d)z level.}
    \label{tab:energetics}
    \begin{tabular}{lcccc}
   \toprule
    Reaction & Label & $\Delta H^{ \textrm{R}}$ / $\Delta H^{ \textrm{R}}_{\textrm{CPCM}}$ &
    $\Delta H^{ \ddagger}$ / $\Delta H^{ \ddagger}_{\textrm{CPCM}}$ & $\Omega$ / $T_{\textrm{c}}$ \\
   \bottomrule
    \ce{CH3CHO + H -> CH3CO + H2}   & \ref{eq:h1} & -63.1 / -55.7 & 18.5 / 23.9 & 1450$i$ / 332 \\
    \ce{CH3CHO + H -> CH3CH2O}      & \ref{eq:c2} & -60.3 / -54.2 & 31.0 / 31.2 & 1260$i$ / 289 \\
    \ce{CH3CHO + H -> CH2CHO + H2}  & \ref{eq:h3} & -32.7 / -31.5 & 41.2 / 40.8 & 1644$i$ / 377 \\
    \ce{CH3CHO + H -> CH3CHOH}      & \ref{eq:o4} & -101.3 / -101.1 & 47.3 / 51.2 & 2005$i$ / 459 \\
    \ce{CH3CHO + H -> CH4 + HCO}    & \ref{eq:c5} & -86.6  / -76.7  & 142.6 / 145.4 & 1597$i$ / 366 \\
    \bottomrule
   \end{tabular}
   \tablefoot{$\Delta H^{ \textrm{R}}$ and $\Delta H^{ \ddagger}$ are determined from a preconverged pre-reactant complex (PRC) as $\Delta H^{ \textrm{R}, \ddagger}$ = $H_{\rm Prod, TS}$ - $H_{\rm PRC}$. \rev{ZPVE are determined at the rev-DSD-PBEP86(D4)/jun-cc-pV(T+d)Z level}}
\end{center}
\end{table*}

\begin{figure}[h]
        \centering
        \vspace{1.5em}
        \includegraphics[width=\linewidth]{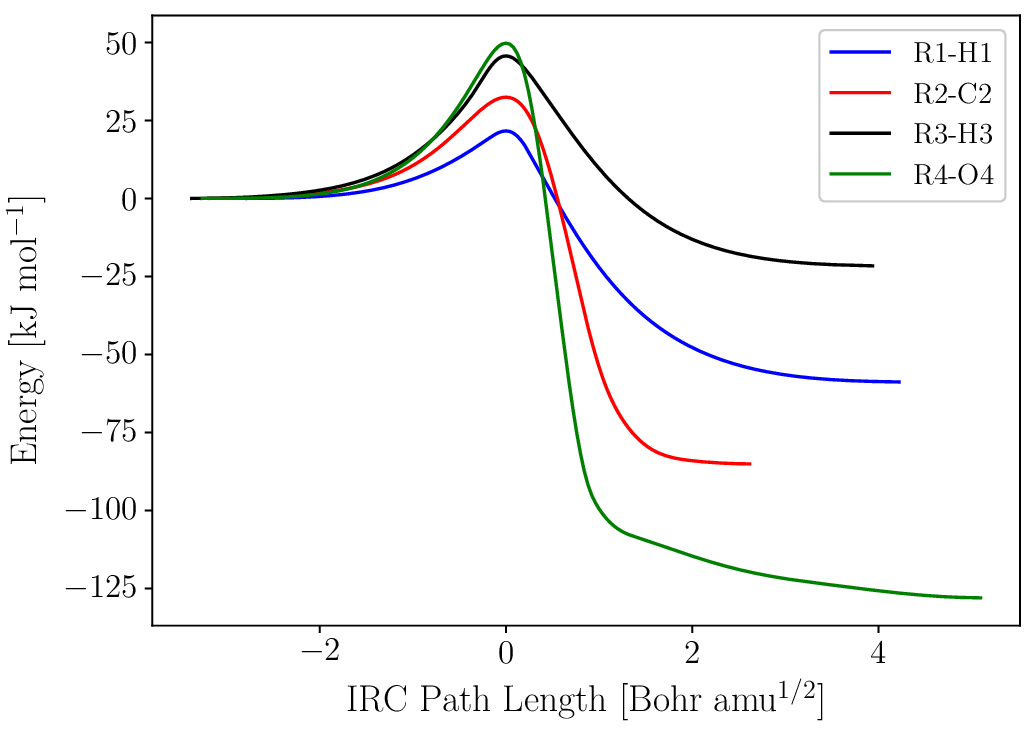} \\
        \caption{IRC profiles for reactions \ref{eq:h1}--\ref{eq:o4} (We omit reaction \ref{eq:c5} due to the high $\Delta H^{ \ddagger}$). \rev{IRC profiles are presented at the rev-DSD-PBEP86(D4)/jun-cc-pV(T+d)Z not corrected by ZPVE, double-level CCSD(T), or implicit water ice environment (CPCM).} }
        \label{fig:theo2}
\end{figure} 

The energetic parameters for the reactions, reaction energies ($\Delta H^{ \textrm{R}}$), activation energies ($\Delta H^{ \ddagger}$) and imaginary frequencies of vibration at the TS ($\Omega$) are gathered in Table \ref{tab:energetics}. In the first place, we confirmed that all possible reactions are exothermic reactions that proceed with a moderate $\Delta H^{ \ddagger}$, as is expected for closed-shell organic molecules. It is evident from the data in Table \ref{tab:energetics} that reaction \ref{eq:c5} is, under no circumstances, competitive with reactions \ref{eq:h1}--\ref{eq:o4}, due to the extremely high activation energies. This conclusion allows us to determine that all \ce{CH4} and their isotopologues are formed in the experiment from radical-radical recombinations from secondary products of the reaction (See Section \ref{sec:merging}). The other reactions present barriers between $\sim$20--50 kJ mol$^{-1}$ with varying values of $\Omega$. For example, the reactions with the two lowest $\Delta H^{\ddagger}$ also show the lowest $\Omega$. To be completely certain of the dominant reaction product, a kinetic analysis is required. Despite such an analysis being carried out in Section \ref{sec:rates}, a qualitative assessment of the influence of quantum tunneling can be done from the reaction coordinate (IRC) profiles. Such IRC profiles can be visualized in Figure \ref{fig:theo2}. From them, we observe that both reactions \ref{eq:h1} and \ref{eq:h3} have wider barriers than reactions \ref{eq:c2} and \ref{eq:o4}. Besides, both H-addition reactions, \ref{eq:c2} and \ref{eq:o4}, are also more exothermic, which has also an impact on the tunneling probability. Overall, based on the energetics and $T_{\textrm{c}}$ (Equation \ref{eq:cooling}) it is not possible to determine which is the dominant product of the reaction. Kinetic simulations such as the ones in Section \ref{sec:rates} are necessary to discriminate them.

\subsubsection{Kinetic Analysis and reactions with Deuterium} \label{sec:rates}

\begin{figure}[h]
    \centering
    \vspace{1.5em}
    \includegraphics[width=\linewidth]{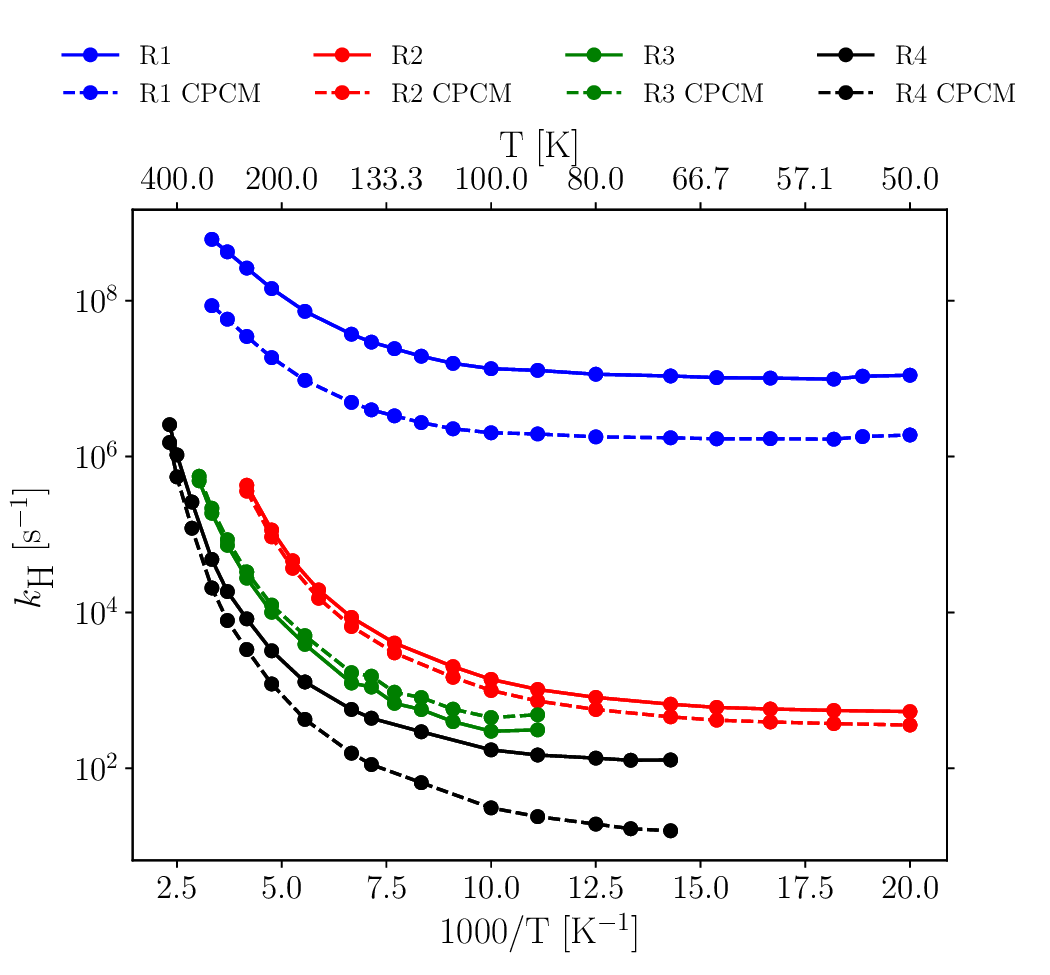} \\
    \caption{Arrhenius plot of the hydrogenation rate constants ($k_{\textrm{H}}$) for reactions \ref{eq:h1}--\ref{eq:o4} including (dashed lines) and without including (solid lines) an implicit solvation model. All rate constants are corrected for implicit surface. Rate constants are presented until lowest temperature achievable for each instanton in the sequential cooling scheme. }
    \label{fig:theo3}
\end{figure} 

We begin our instanton calculations using the sequential cooling scheme at a temperature of approximately $0.7 T_{\textrm{c}}$, starting from a closed instanton path consisting of 36 beads. As described in Section \ref{sec:methods}, we progressively converge the instanton paths at lower temperatures. To ensure convergence with respect to the number of beads at low temperatures, we increase the number of beads to 70 for temperatures below approximately 140 K in all reactions. The hydrogenation rate constants, $k_{\textrm{H}}$, are presented in Figure \ref{fig:theo3}. We omit the calculation of rate constants for Reaction \ref{eq:c5} because instanton calculations are computationally demanding, and this reaction has too high an activation energy to be competitive. \rev{The rate constants are computed down to different lower temperature limits for each reaction.} This variation arises because the convergence of instanton paths depends on the reaction, with the abstraction on the \ce{CH3} moiety being particularly challenging to converge. However, in the cases we studied, the rate constants are converged or close to the asymptotic tunneling limit, so $k_{\textrm{H}}(T_{\textrm{last}}) \approx k_{\textrm{H}}(10)$. Since the reaction outcome is dominated by Reaction \ref{eq:h1}, the lack of rate constants at temperatures below 50 K for Reactions \ref{eq:h3} or \ref{eq:o4} does not affect our conclusions. It is worth noting that the effect of the implicit solvation scheme varies among different reactions, with Reactions \ref{eq:h1} and \ref{eq:o4} showing the most significant changes. Specifically, at low temperatures, these reactions are slower when considering an implicit water matrix due to the increase in $\Delta H^{\ddagger}_{\textrm{CPCM}}$ found for the reactions. Nevertheless, this increase does not alter the overall kinetics of the system regarding competitive reaction pathways.

A quick inspection of Figure \ref{fig:theo3} reveals that Reaction \ref{eq:h1} is the fastest across all temperatures, consistently dominating over all other reaction pathways. Furthermore, aside from Reaction \ref{eq:h1}, all other reactions exhibit rate constants that are competitive with hydrogen atom diffusion \citep{asg17,SENEVIRATHNE201759}, with diffusion rates ($k_{\textrm{Diff}}$) in typical potential sites ranging from $10^{3}$ to $10^{5}$ s$^{-1}$ at 10 K \citep[see, for example, Figure 7, bottom panel of][]{SENEVIRATHNE201759}. This competition suggests that even if a pre-reactive complex favoring reactions other than Reaction \ref{eq:h1} is formed—perhaps due to a favorable orientation of the radical attack—diffusion away from the reaction site in a reaction-diffusion competition \citep{changGasgrainChemistryCold2007} will either completely or partially suppress these alternative reaction pathways. Consequently, we conclude that Reaction \ref{eq:h1} is the sole outcome of the \ce{CH3CHO + H} reaction. This conclusion allows us to rationalize the results obtained in our experiments (see Section \ref{sec:ch3o}).

\begin{figure}[h]
    \centering
    \vspace{1.5em}
    \includegraphics[width=\linewidth]{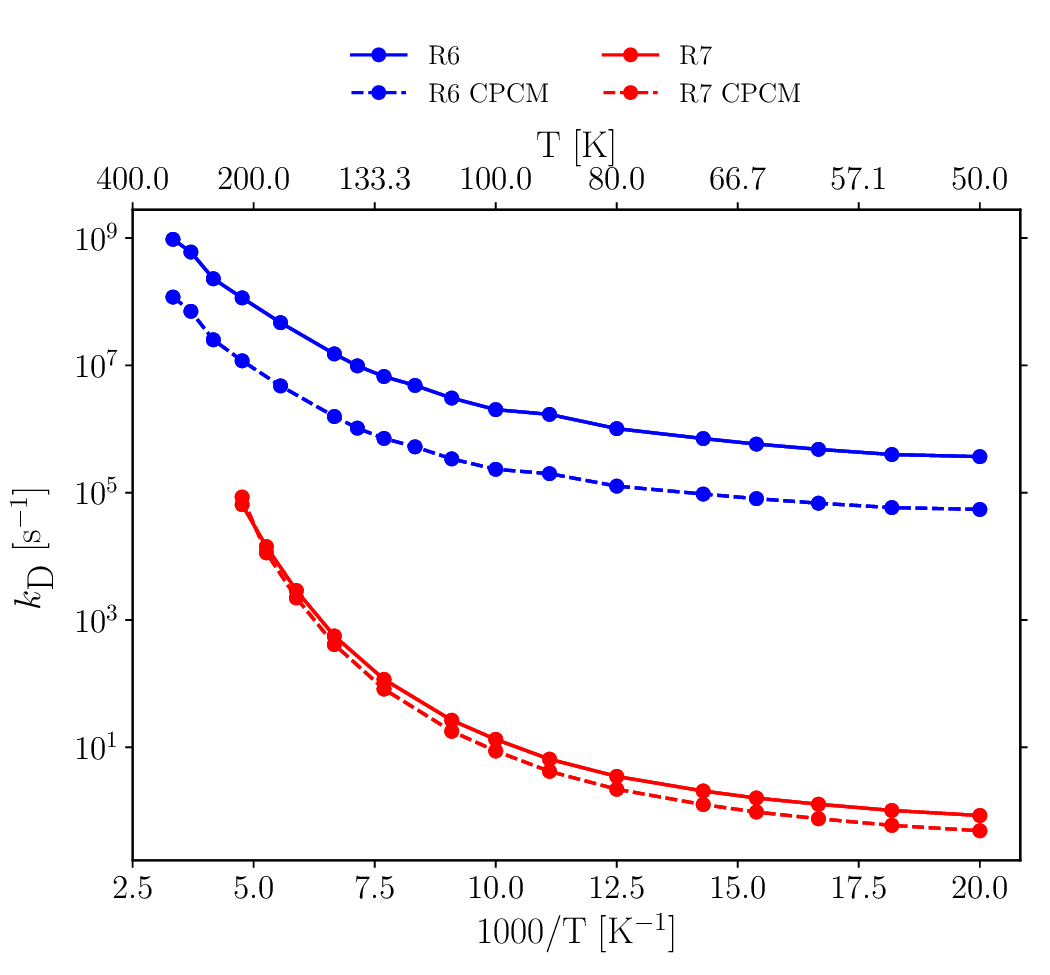} \\
    \caption{Similar to Figure \ref{fig:theo3}  but including deuterium as incoming particle ($k_{\textrm{D}}$) for reactions \ref{eq:h1d} and \ref{eq:c2d}}
    \label{fig:theo4}
\end{figure} 

We also examined deuteration rate constants, $k_{\textrm{D}}$. For these calculations, we focused on the H1 and C2 positions. For example:

\begin{chequation}
    \begin{align}
        \ce{CH3CHO + D &-> CH3CO +HD} \label{eq:h1d} \\
        \ce{CH3CHO + D &-> CH3CHDO } \label{eq:c2d}
    \end{align}
\end{chequation}

\noindent because they have the lowest $\Delta H^{ \ddagger}$ and the highest $k_{\textrm{H}}$. The rate constants are shown in Figure \ref{fig:theo4}. In both cases we observe a reduction of $k_{\textrm{D}}$ over $k_{\textrm{H}}$ due to the reduced efficiency of quantum tunneling. The kinetic isotope effects (KIE; calculated as $k_{\textrm{H}}$ / $k_{\textrm{D}}$) at 50 K\footnote{\rev{While we formally calculate $k_{\textrm{H,D}}$ down to 50 K, a extrapolation to 10 K would not change the KIE significantly as \ref{eq:h1}, \ref{eq:h1d}, \ref{eq:c2} and \ref{eq:c2d} are almost at the horizontal asymptote characteristic of unimolecular deep tunneling (Figures \ref{fig:theo3} and \ref{fig:theo4}).}} that we observe according to our calculations are, \rev{$\textrm{KIE}_{\textrm{\ref{eq:h1}/\ref{eq:h1d}}}$=34. 7 and $\textrm{KIE}_{\textrm{\ref{eq:c2}/\ref{eq:c2d}}}$=734.1}, calculated including the implicit solvent approach. The KIE is larger in the case of reaction \ref{eq:c2d}, because this reaction is a D addition and quantum tunneling is reduced, compared to reaction \ref{eq:h1d}, which is an H abstraction and has a higher tunneling probability.

\subsubsection{Hydrogenation of the \ce{CH3CO} radical} \label{sec:ch3o}

The reaction \ce{CH3CHO + H} leads almost exclusively to \ce{CH3CO}. Therefore, the \ce{CH3CHO} reaction network must be completed by relying on the chemistry of such a radical. In this section we study the hydrogenation of \ce{CH3CO} following a scheme similar to that shown in Figure \ref{fig:theo1} (only this time without the H1 position). The study of the \ce{CH3O + H} reaction has the peculiarity that it should be studied in the open-shell singlet channel, i.e. using a biradical wavefunction. Radical-radical reactions generally show lower barriers than the radical-closed shell reactions, so we first check which of the following reactions are barrierless using PES scans:

\begin{chequation}
\begin{align}
   \ce{CH3CO + H &-> CH3CHO} \label{eq:ac-c2} \\
   \ce{CH3CO + H &-> H2CCO + H2} \label{eq:ac-h3} \\
   \ce{CH3CO + H &-> CH3COH} \label{eq:ac-o4} \\
   \ce{CH3CO + H &-> CH4 + CO} \label{eq:ac-c5} 
\end{align}
\end{chequation}

\begin{figure}[h]
   \centering
   \vspace{1.5em}
   \includegraphics[width=\linewidth]{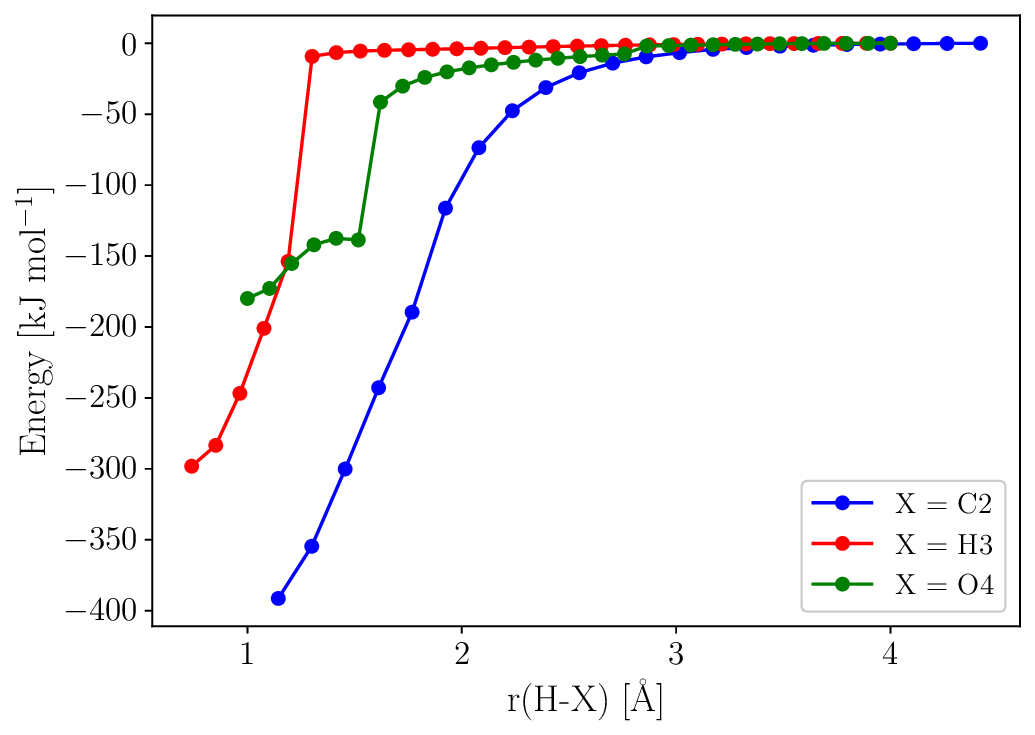} \\
   \caption{Potential energy scan for \rev{Reactions \ref{eq:ac-c2} (blue curve), \ref{eq:ac-h3} (red curve) and \ref{eq:ac-o4} (green curve)} at the revDSD-PBEP86(D4)/jun-cc-pV(T+d)z level using a broken symmetry formalism. The discontinuities in r(H-H3) and r(H-O4) are caused by insufficient resolution of the PES scan. The small bump in r(H-O4) is attributted to the constraints in the C-H bonds required to carry out the scan (See text).}
   \label{fig:ac-theo1}
\end{figure} 

\begin{table*}[bt!]
   \begin{center}
    \caption{Similar to Table \ref{tab:energetics}, but for the \ce{CH3CO + H} reaction. We added a column $k_{\textrm{Eckart}}^{50}$ denoting the rate constant at 50 K (s $^{-1}$) computed using an asymmetric Eckart fit to the barrier in reaction \ref{eq:ac-c5} \rev{with/without CPCM correction}. \rev{All} values are reported at the revDSD-PBEP86(D4)/jun-cc-pV(T+d)z level with and without including implicit solvent corrections. The A(B) notation symbolizes A$\times$10$^{B}$.}
    \label{tab:ac-energetics} 
    \begin{tabular}{lccccc}
   \toprule
   Reaction & Label & $\Delta H^{ \textrm{R}}$ / $\Delta H^{ \textrm{R}}_{\textrm{CPCM}}$ &
   $\Delta H^{ \ddagger}$ / $\Delta H^{ \ddagger}_{\textrm{CPCM}}$ & $\Omega$ & $k_{\textrm{Eckart}}^{50}$ / $k_{\textrm{Eckart, CPCM}}^{50}$\\
   \bottomrule
    \ce{CH3CO + H -> CH3CHO}       & \ref{eq:ac-c2} & -398.9 / -406.1 $^{a}$ & N/A$^{b}$  & N/A$^{b}$ & N/A$^{b}$ \\
    \ce{CH3CO + H -> H2CCO + H2}  & \ref{eq:ac-h3} &  -298.8 / -296.6 $^{a}$& N/A$^{b}$ & N/A$^{b}$ & N/A$^{b}$ \\
    \ce{CH3CO + H -> CH3COH}      & \ref{eq:ac-o4} & -154.9 / -179.1 $^{a}$& N/A$^{b}$ & N/A$^{b}$ & N/A$^{b}$ \\
    \ce{CH3CO + H -> CH4 + CO}    & \ref{eq:ac-c5} & -395.3 / -387.9 & 13.2 / 16.2 & 928$i$ & 2.0(7) / 1.4(6)$^{c}$ \\
    \bottomrule
   \end{tabular}
   \tablefoot{$^{a}$ - Calculated from the bimolecular system as $\Delta H^{ \textrm{R}}$ = $H_{\rm Prod}$ - ($H_{\ce{CH3CO}}$ + $H_{\ce{H}}$) $^{b}$ - Barrierless. $^{c}$ - Unimolecular rate constants obtained using a reactant state at large \ce{CH3CO} and H distance, as converging a pre-reactant complex was impossible due to the barrierless channels presence nearby. }
\end{center}
\end{table*}

\noindent The scans for the barrierless reactions are shown in Figure \ref{fig:ac-theo1}. Of all the above reactions, only reaction \ref{eq:ac-c5} was not found to be barrierless. All other reactions show a barrierless profile. \rev{We remark the difficulty of obtaining clean scans for these, except for reaction \ref{eq:ac-c2}, because in the latter case there are no competing channels in the direction of the scan.} The other scans have discontinuities that we attribute to a combination of a low resolution and hindered conformational changes from competitive barrierless pathways during the coordinate scan. For example, the scan for reaction \ref{eq:ac-o4} shows a small bump at about 1.4 \AA. This is due to the fact that in this scan we had to constrain the C-H (\ce{CH3} moiety) bond distance to prevent the optimizer from visiting the barrierless reaction \ref{eq:ac-h3} at long distances. This results in a small stiffness of the structure leading to this bump. Our calculations still predict the barrierless reaction. We recall that the calculations involving these reactions are carried out using a broken-symmetry formalism, at the revDSD-PBEP86(D4)/jun-cc-pV(T+d)z level, without correction with coupled-cluster methods.  This, combined with the also approximate nature of the CPCM implicit solvent scheme, makes the second set of calculations less accurate than those shown in section \ref{sec:ch3cho}. The energetic values for the reactions are collected in Table \ref{tab:ac-energetics}. From the table we can see that the $\Delta H^{ \ddagger}$ for reaction \ref{eq:ac-c5} (counterpart of \ref{eq:c5}) is significantly lower in this second set than in the previous reactions for the hydrogenation of acetaldehyde. We also observe the effect of the implicit solvation correction on some energetic parameters of the reaction. The abundant presence of \ce{CH4} and its isotopologues in our experiments supports the reaction \ref{eq:ac-c5}, even with a barrier (see next paragraph and section \ref{sec:experiments:H} and \ref{sec:merging}). Although our results for this reaction are qualitative, they are in agreement with the experiments. However, for a more detailed study of \ref{eq:ac-c5}, we suggest using multireference methods. All other reactions are barrierless. Finally, it is also worth noting that our results in this section are not affected by deuteration, since tunneling does not determine the viability of the reaction (for reactions \ref{eq:ac-c2}-\ref{eq:ac-o4}) or does not contribute significantly (reaction \ref{eq:ac-c5}, which requires the fragmentation of a C-C bond).

The possibility of finding products beyond those obtained in barrierless channels can be easily rationalized if we consider the typical reaction-diffusion competition scheme \citep{changGasgrainChemistryCold2007}. Assuming that the desorption of both reactants and the diffusion of \ce{CH3CHO} are negligible, the rate constant for the reaction under this formalism is \citep{changGasgrainChemistryCold2007}:

\begin{equation} \label{eq:dr}
   k = \dfrac{k_{\textrm{R}}}{k_{\textrm{R}} + k_{\textrm{Diff}}},
\end{equation}

\begin{figure}[h]
   \centering
   \vspace{1.5em}
   \includegraphics[width=0.8\linewidth]{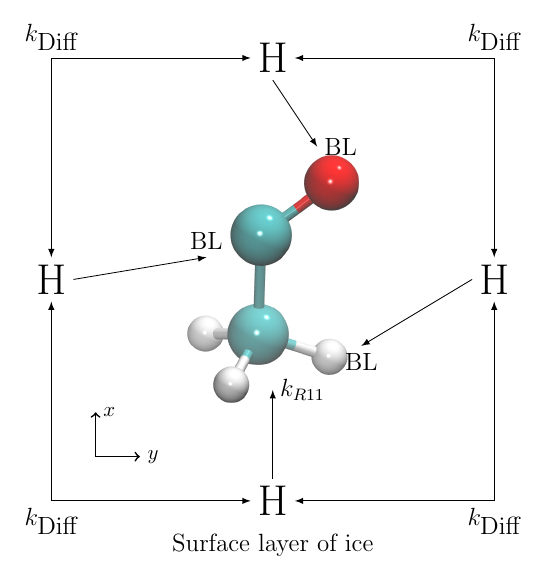} \\
   \caption{Schematic view of the \ce{CH3CO + H} reaction. Depending on the initial configuration of the H atoms, the different $k_{\textrm{R}}$ must compete with diffusion, $k_{\textrm{Diff}}$. If $k_{\textrm{R}}$ $\gg$ $k_{\textrm{Diff}}$, then the reaction would dominate even in the presence of a barrierless channel, as shown in the figure. BL means barrierless. } 
   \label{fig:theo5}
\end{figure} 

\noindent where ``R'' and ``Diff'' represent reaction and diffusion.In the anisotropic potential created by an amorphous polar environment such as ASW, the orientation from which H atoms approach is random. Assuming that all directions are equally possible for reactions \ref{eq:ac-c2}--\ref{eq:ac-c5}, the equation \ref{eq:dr} shows that if each reaction starts from different positions, as is the case, then moving out of the reaction center requires at least one diffusion step. If $k_{\textrm{R}}$ is much larger than $k_{\textrm{Diff}}$, then the reaction can start from that orientation, even if a barrierless channel is available. This is essentially different from gas phase reactivity. A scheme for this explanation is shown in figure \ref{fig:theo5}. The arguments based on reaction-diffusion competition are therefore able to rationalize the experimental results in the light of the theoretical calculations, as we show in Section \ref{sec:merging}. However, not all reaction orientations are equally weighted, resulting in different branching ratios for reactions. Determining the exact branching ratios for these barrierless reactions from theoretical calculations is challenging.The same arguments presented in this section apply to the reactions studied in section \ref{sec:ch3cho} (reactions \ref{eq:h1}--\ref{eq:c5}), but in this case reactions other than reaction \ref{eq:h1} are competitive with diffusion, $k_{\textrm{R}}$$\sim$$k_{\textrm{Diff}}$, and therefore reaction \ref{eq:h1} is dominant.  

Our theoretical results for the hydrogenation of \ce{CH3CO} are in reasonable agreement with those presented in \citet{ibrahim_formation_2022}, although several differences appear. In their study, \citet{ibrahim_formation_2022} indicate that the reaction \ref{eq:ac-c5} is barrierless, while the reaction \ref{eq:ac-h3} has an activation energy of $\sim$ 18 kJ mol$^{-1}$.  This tendency contradicts our calculations and experiments. However, this discrepancy does not change the picture of both works, except for the absence of \ce{H2CCO} by reaction \ref{eq:ac-h3}, which we found very fast, and which is supported by experiments, since all other reactions remain very fast in both our works. We believe that the reason for the discrepancy could be a non-biradical wavefunction in the case of \citet{ibrahim_formation_2022}. Starting our calculations from an ionic wavefunction, we can reproduce the results of \citet{ibrahim_formation_2022} for the reaction \ref{eq:ac-h3} with even higher activation energies ($\sim$ 30 kJ mol$^{-1}$). 

\subsection{Experiments} \label{sec:experiments}

\subsubsection{Hydrogenation experiments} \label{sec:experiments:H}

\begin{figure}[ht]
    \centering
    \includegraphics[width=\linewidth]{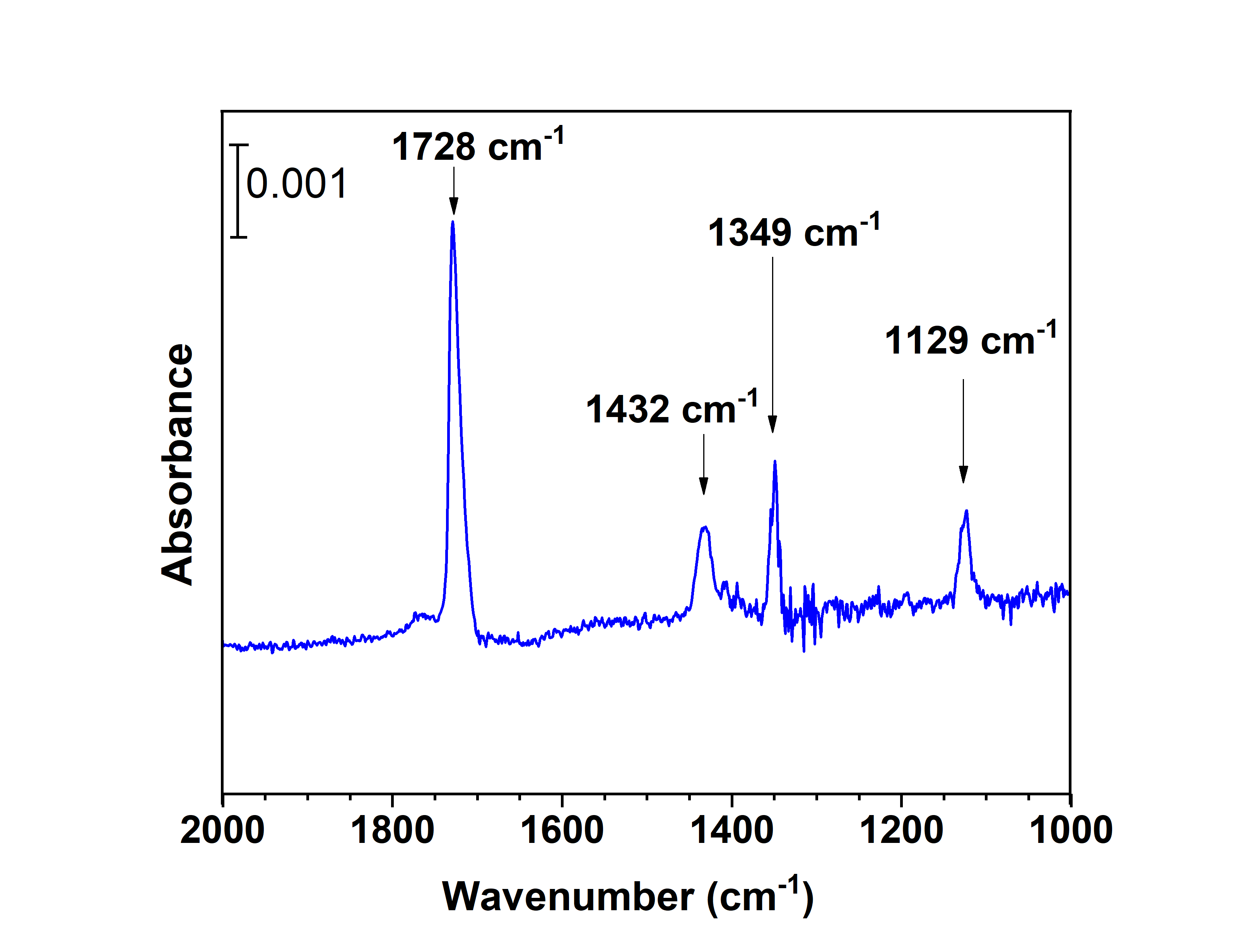}
    \caption{Fourier-transform infrared profile of \ce{CH3CHO} on c-ASW at 10~K}
    \label{fig:CH3CHOIR}
\end{figure}

Figure \ref{fig:CH3CHOIR} shows an FTIR spectrum of the solid \ce{CH3CHO} on c-ASW at 10~K. The most intense absorption peak at 1728 cm$^{-1}$ was attributed to the C-O stretching band ($\nu$$_s$C=O) in \ce{CH3CHO}. Additionally, other infrared bands were observed at 1432, 1349, and 1124 cm$^{-1}$ can be assigned to the CH$_2$ scissors, CH$_3$ symmetric deformation, and C-C stretch of the solid \ce{CH3CHO}, respectively \citep{Bennet2005,Bisschop2007}. 

\begin{figure}[ht]
   \centering
   \includegraphics[width = \linewidth]{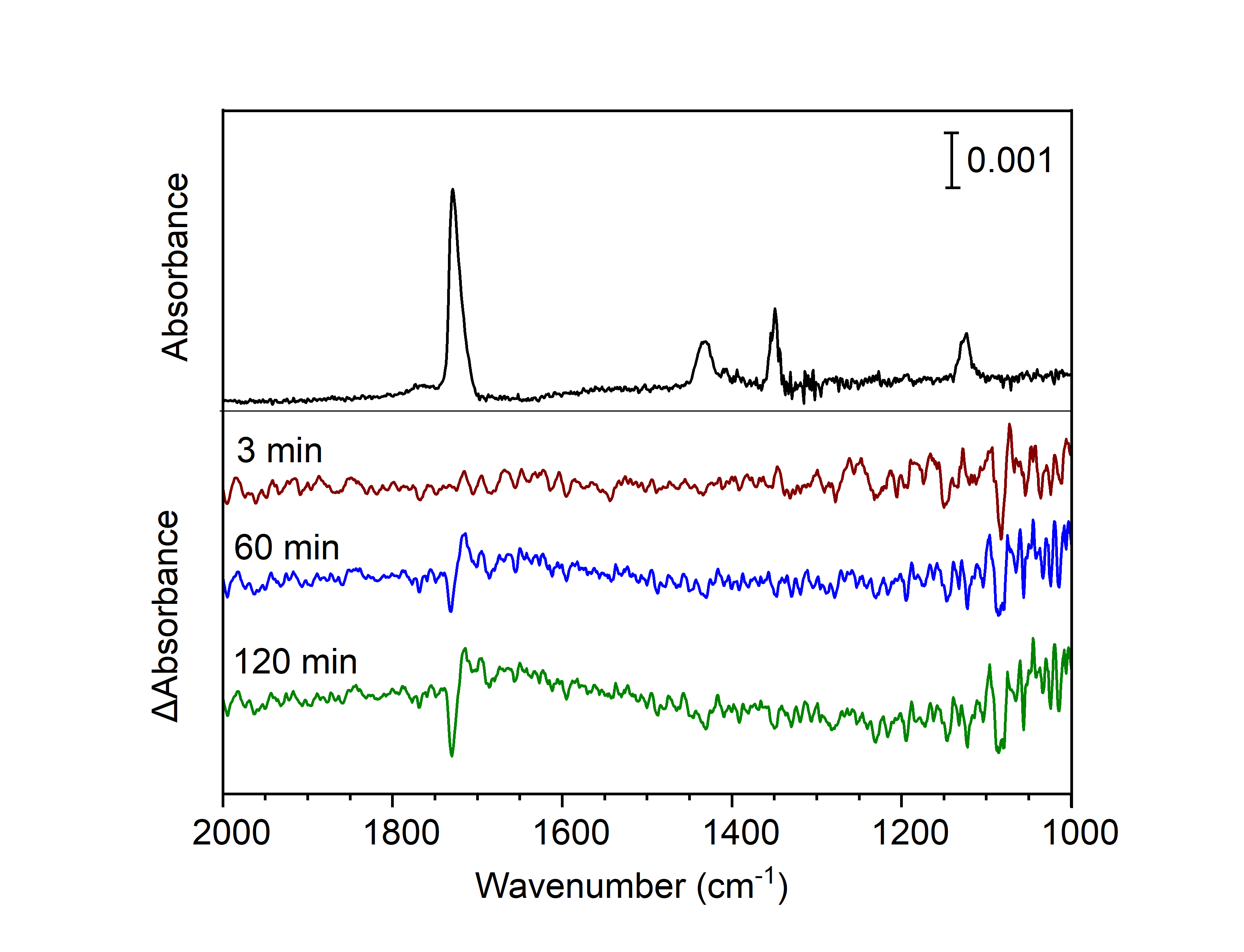}
   \includegraphics[width = \linewidth]{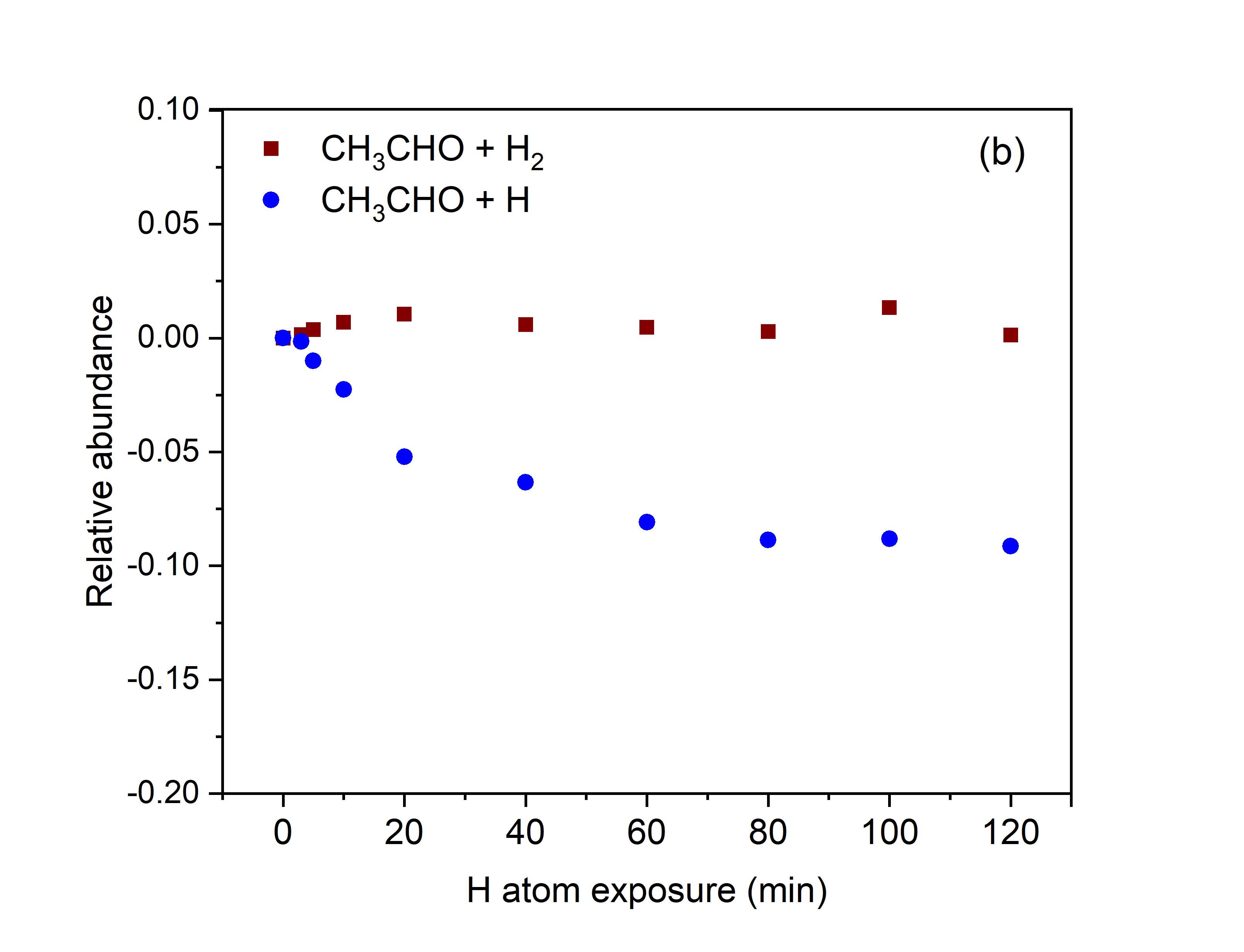}
   \caption{(a) Variation in the difference spectra of the solid \ce{CH3CHO} after exposure to H atoms for 3, 60, and 120 min at 10~K. (b) Relative abundance of \ce{CH3CHO} following a 2-hour exposure to H atoms (blue circles) compared to that with H$_2$ molecules (red squares) on c-ASW at 10~K}
   \label{fig:variationinFTIRspectraCH3CHO}
\end{figure}

Figure \ref{fig:variationinFTIRspectraCH3CHO}(a) shows the difference spectrum of solid \ce{CH3CHO} after exposure to H atoms for up to 2 hours compared to the initial spectrum of unexposed \ce{CH3CHO}. The intensity of the C-O stretching band at 1728 cm$^{-1}$ decreased with atom exposure time, indicating the loss of \ce{CH3CHO}.  Figure \ref{fig:variationinFTIRspectraCH3CHO}(b) shows variations in the relative abundance of \ce{CH3CHO} after exposure to H atoms at 10~K with relevance to atom exposure times, with a blank experiment for \ce{H2} exposure. The consumption of \ce{CH3CHO} was estimated to be only 10$\%$ relative to the initial amount of \ce{CH3CHO} after exposure to H atoms for up to 2 hours, while no decrease was observed after exposure to \ce{H2} molecules. We expect the formed \ce{CH3CO} radicals from reaction \ref{eq:h1} to further react with H atoms on the surface according to the reactions (\ref{eq:ac-c2} - \ref{eq:ac-c5}). Unfortunately, we could not observe any distinct infrared features for products from the pre-deposition of \ce{CH3CHO} and H atoms on c-ASW at 10~K, probably because of the small amount of products. The loss of \ce{CH3CHO} may be partially due to reactive desorption. We cannot quantitatively evaluate this in any of our experiments, even considering the low desorption energy of \ce{CH3CHO} of $\sim$3500 K \citep{molpeceres_desorption_2022,ferrero_acetaldehyde_2022}, something we discuss in more detail in section \ref{sec:implications}.

\begin{figure}[ht!]
   \centering
   \includegraphics[width = \linewidth]{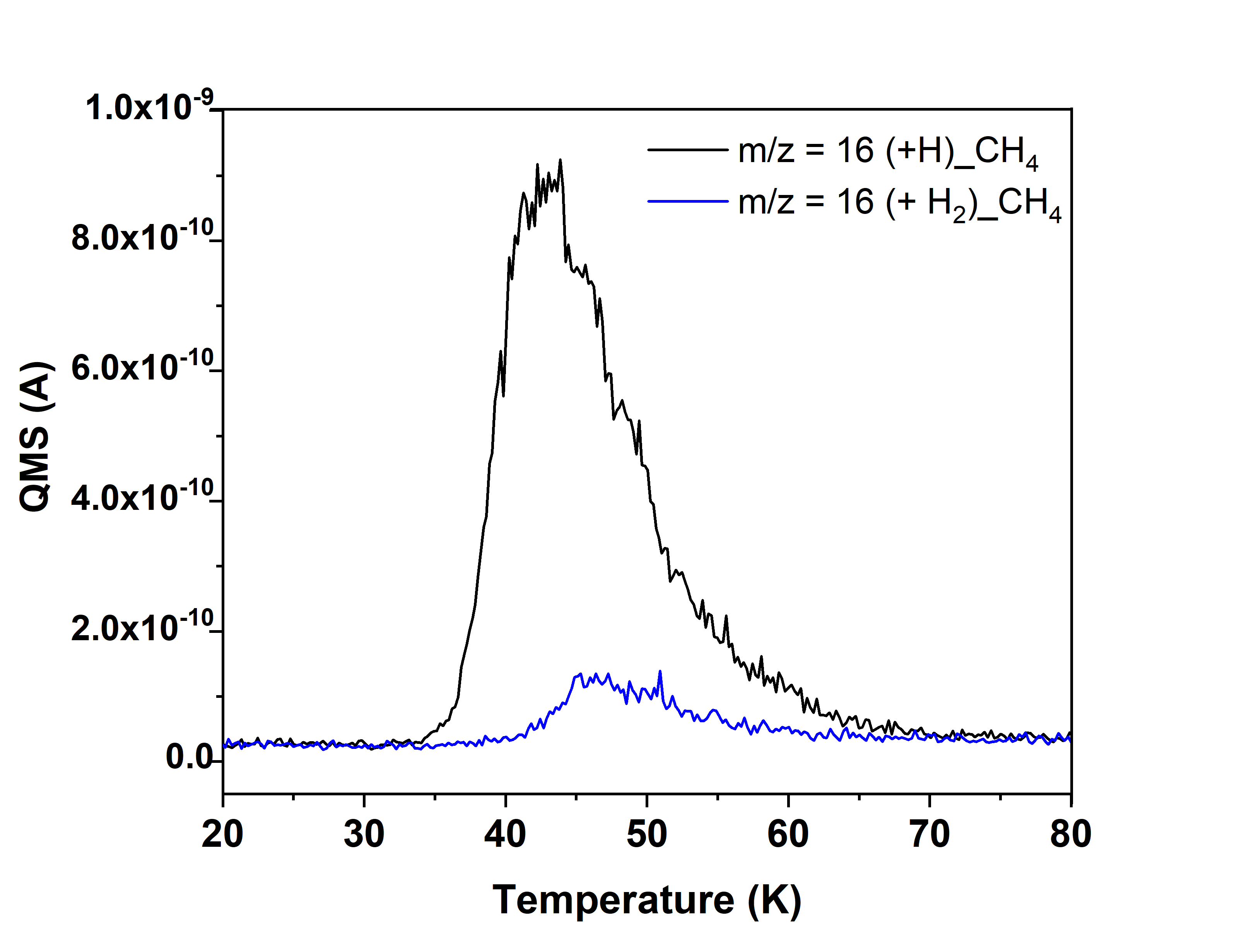}
   \includegraphics[width = \linewidth]{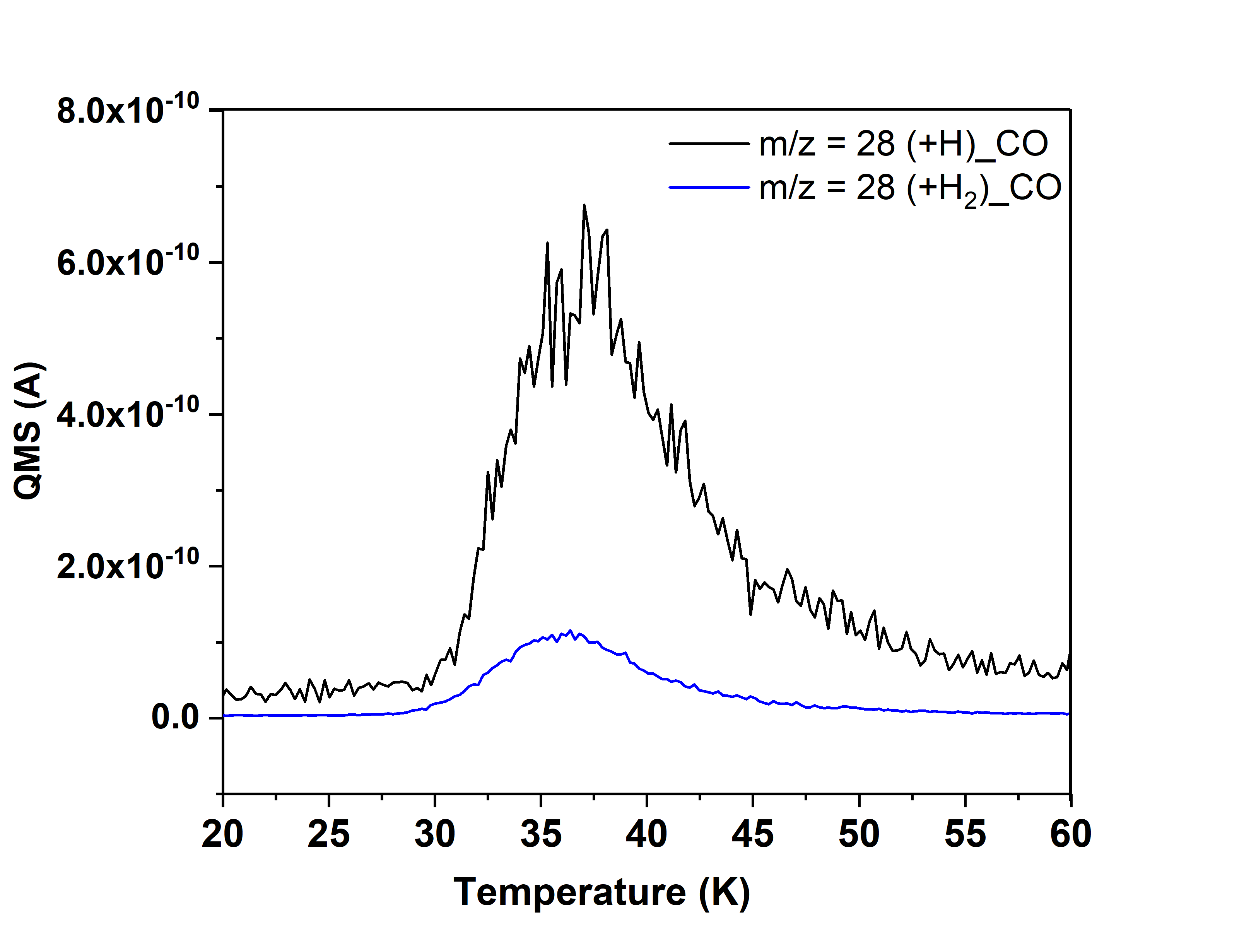}
   \includegraphics[width = \linewidth]{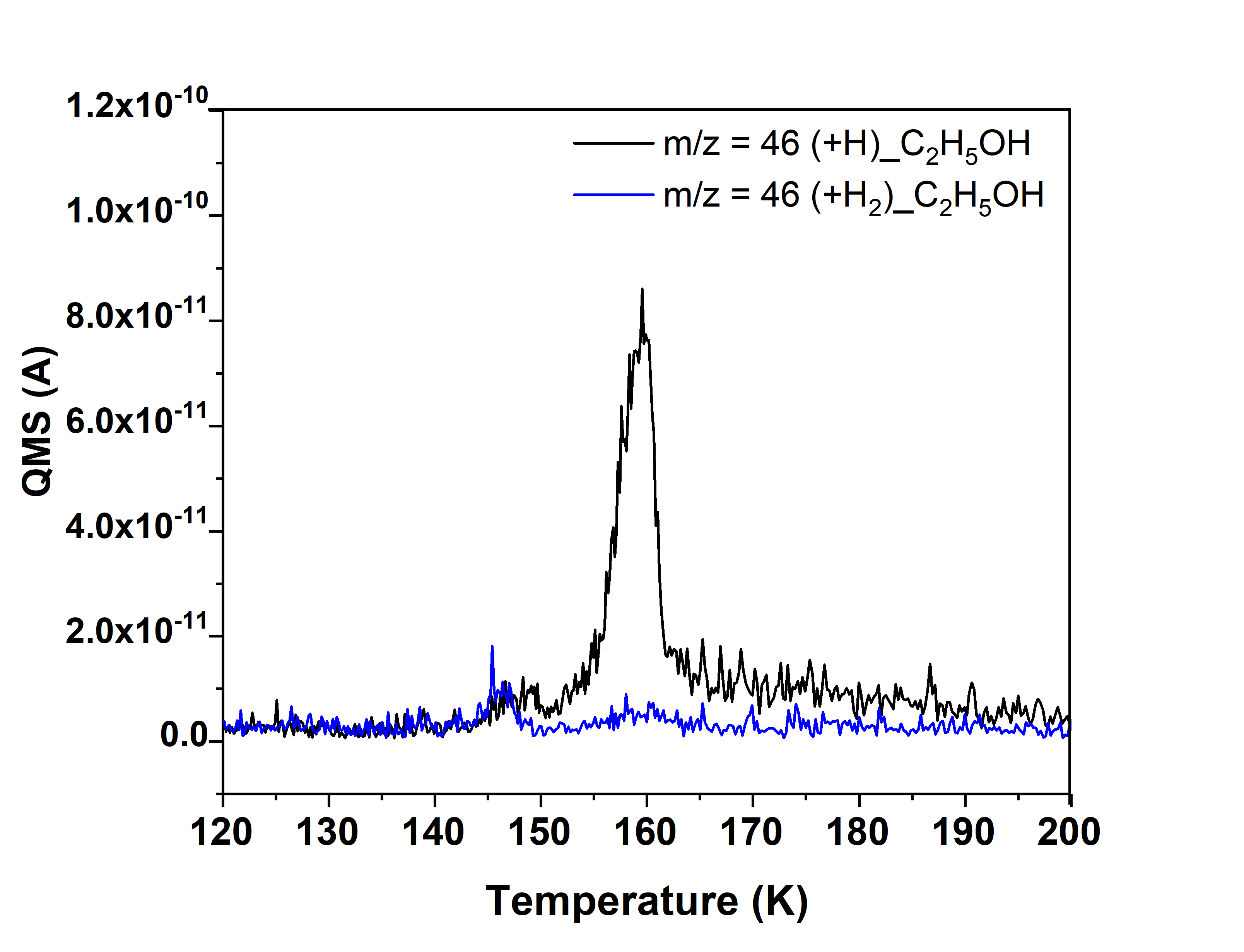}
   \caption{TPD-QMS profile of pre-deposited \ce{CH3CHO} with H atoms (black line) comparable with H$_2$ molecules (blue line) on c-ASW at 10~K: m/z = 16 (top panel), m/z = 28 (middle panel), and m/z = 46 (bottom panel).}
   \label{fig:TPD-QMS_for_predeposition}
\end{figure}

The TPD experiments demonstrated the formation of \ce{CH4} and CO based on the desorption peak observed at m/z~=~16 and m/z~=~28 in the temperature range of 30-60~K and 25-35~K, respectively (Figure \ref{fig:TPD-QMS_for_predeposition}, top and middle panels). These results are in agreement with the computational predictions for the reaction (\ref{eq:ac-c5}) (see Section \ref{sec:ch3o}) and \citep{ibrahim_formation_2022} experiments. In addition, we observed a desorption peak at m/z = 46 in the desorption temperature range 150 - 165~K (Figure \ref{fig:TPD-QMS_for_predeposition}, bottom panel), suggesting the formation of ethanol (\ce{C2H5OH}). This ethanol should be formed by the H addition reaction of \ce{CH3COH}, which was obtained from the reaction \ref{eq:ac-o4}, followed by:

\begin{chequation}
\begin{align}
   \ce{CH3COH + H &-> CH3CHOH} \\
   \ce{CH3CHOH + H &-> C2H5OH}
\end{align}
\end{chequation}

\begin{figure}[ht]
    \centering
    \includegraphics[width=\linewidth]{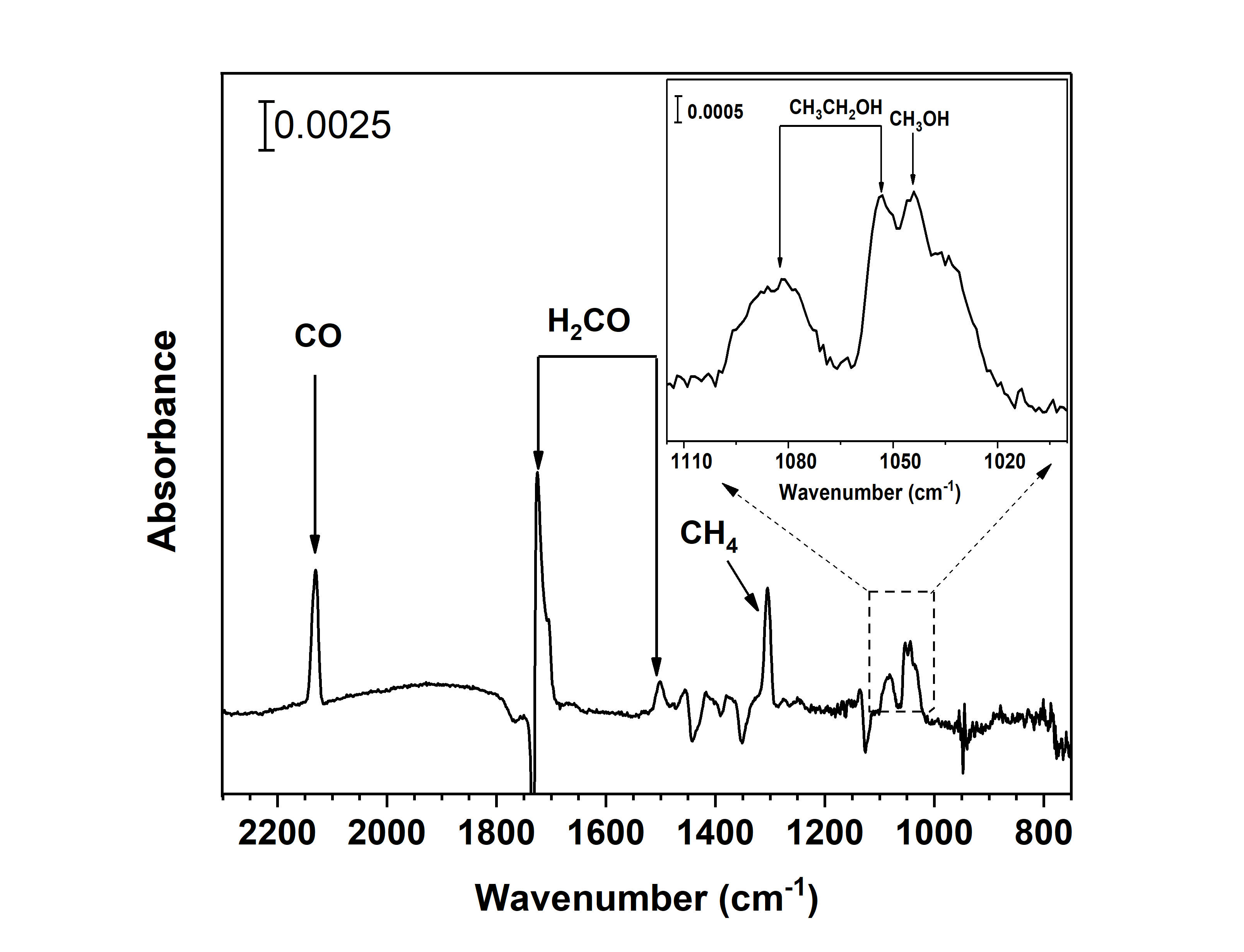}
    \caption{FTIR spectrum of codeposition of \ce{CH3CHO} and H atoms on Al at 10~K.}
    \label{fig:codepositionofCH3CHOandHatoms}
\end{figure}

No distinct IR features of the products of the reaction between \ce{CH3CHO} and H atoms were observed in the pre-deposition experiments, probably due to the small amount of solid \ce{CH3CHO} (1~ML). In order to identify the products of the reaction between \ce{CH3CHO} and H atoms in more detail by FTIR, we performed an additional experiment in which 30 ML of \ce{CH3CHO} was co-deposited with H atoms on the Al substrate at 10~K, a technique suitable for identifying reaction products. Figure \ref{fig:codepositionofCH3CHOandHatoms} shows an FTIR spectrum obtained after co-deposition of \ce{CH3CHO} with H atoms on Al substrate at 10~K. In contrast to the behavior of the pre-deposited \ce{CH3CHO} with H atoms (Figure \ref{fig:variationinFTIRspectraCH3CHO}a), various peaks appeared after the co-deposition of \ce{CH3CHO} with H atoms. The appearance of a sharp peak at 2133 cm$^{-1}$ was assigned to CO \citep{Gerakines1995}. After its formation on the surface, CO should be further hydrogenated to give formaldehyde (H$_2$CO) and methanol (CH$_3$OH), with peaks at 1724 and 1500 cm$^{-1}$ for H$_2$CO \citep{, Hidaka2004} and 1044 cm$^{-1}$ for CH$_3$OH \citep{Nagaoka2007} as observed in the spectrum (Figure \ref{fig:codepositionofCH3CHOandHatoms}). A strong peak observed at 1305 cm$^{-1}$ was attributed to CH$_4$, which was in good agreement with the C-H stretching band of CH$_4$ \citep{Qasim2020}. In addition, two small peaks observed at 1081 and 1050 cm$^{-1}$ were attributed to the CCO asymmetric bending and C-O stretching bands of CH$_3$CH$_2$OH, respectively \citep{Hudson2017}.

\begin{figure}[ht]
   \centering
   \includegraphics[width=\linewidth]{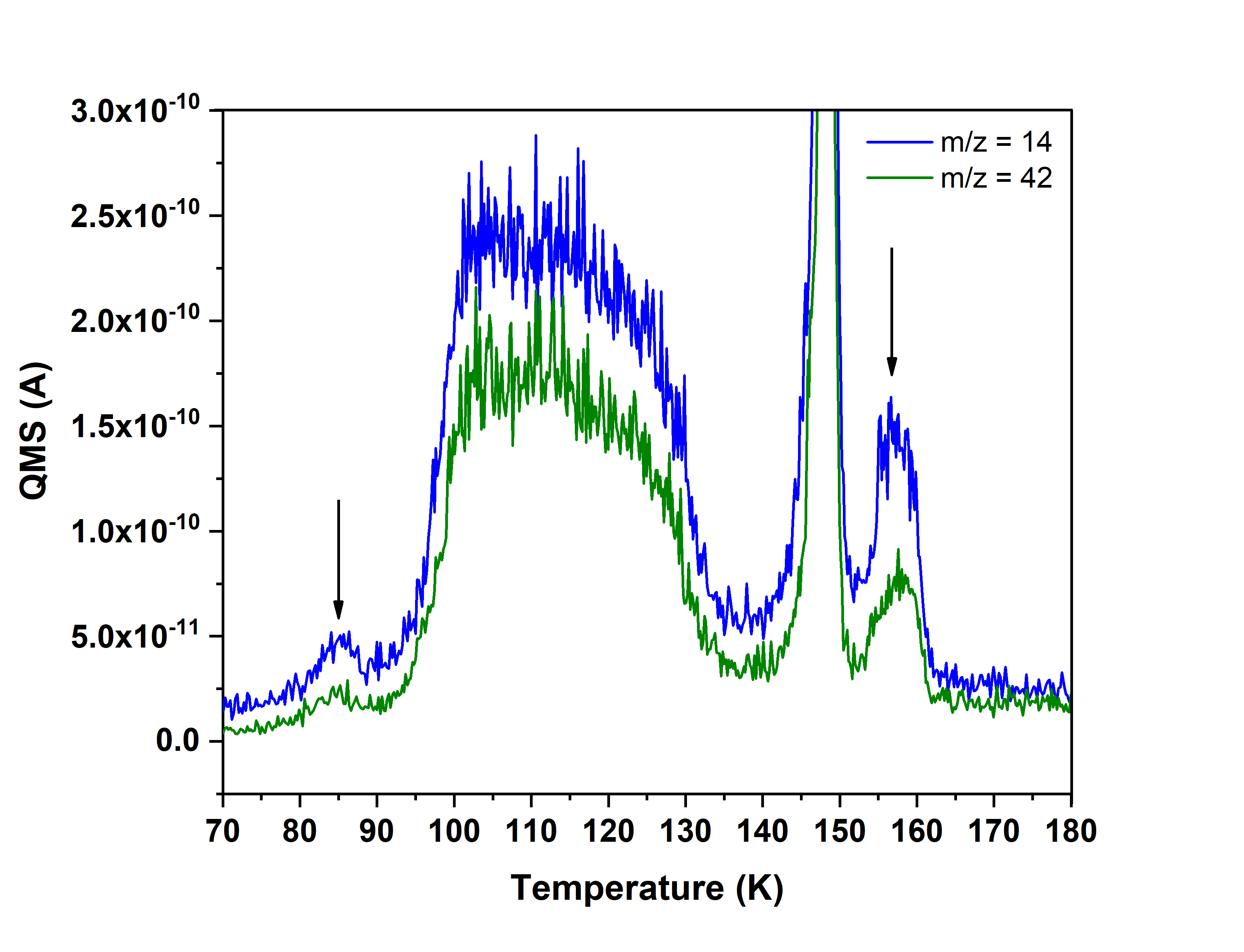}
   \caption{TPD-QMS profile for the m/z = 14 and 42. \rev{Black arrows correspond to features attributable to \ce{H2CCO}}, obtained after the reaction of the pre-deposited \ce{CH3CHO} with H atoms on c-ASW at 10K. The desorption peaks at 100--130 K and 140--150 K are derived from the remaining \ce{CH3CHO} on c-ASW after exposure to H atoms. The peaks at 75--90 K and 150--165 K represent the desorption of \ce{H2CCO} \citep{maity_formation_2015,Chuang2020,Fedoseev2022}.}
   \label{fig:keteneevidence}
\end{figure}

Finally, the presence of ketene (\ce{H2CCO}) was also observed in our co-deposition experiments. We could not detect it in the IR because \ce{H2CCO} is easily hydrogenated \Citep{ferrero_formation_2023}, and its characteristic vibrational frequency (at 2133 cm$^{-1}$) overlaps with the absorption of CO. However, clear evidence for the presence of \ce{H2CCO} in the reaction mixture was observed in our TPD experiments. The desorption peaks corresponding to m/z = 42 and 14 at about 75--90 K and 150--165 K (Figure \ref{fig:keteneevidence}) can be derived from \ce{H2CCO} itself and its fragment (i.e., \ce{CH2}), respectively. The detection of these m/z = 14 (\ce{CH2}) and 42 (\ce{H2CCO}) features in the TPD experiments, consistent with previous studies \citep{maity_formation_2015,Chuang2020,Fedoseev2022}, provided strong evidence that \ce{H2CCO} was formed under the current experimental conditions, as supported by our theoretical models.
  
\subsubsection{Isotopic labeling} \label{sec:results_exp_iso}

\begin{figure}[ht]
   \centering
   \includegraphics[width = \linewidth]{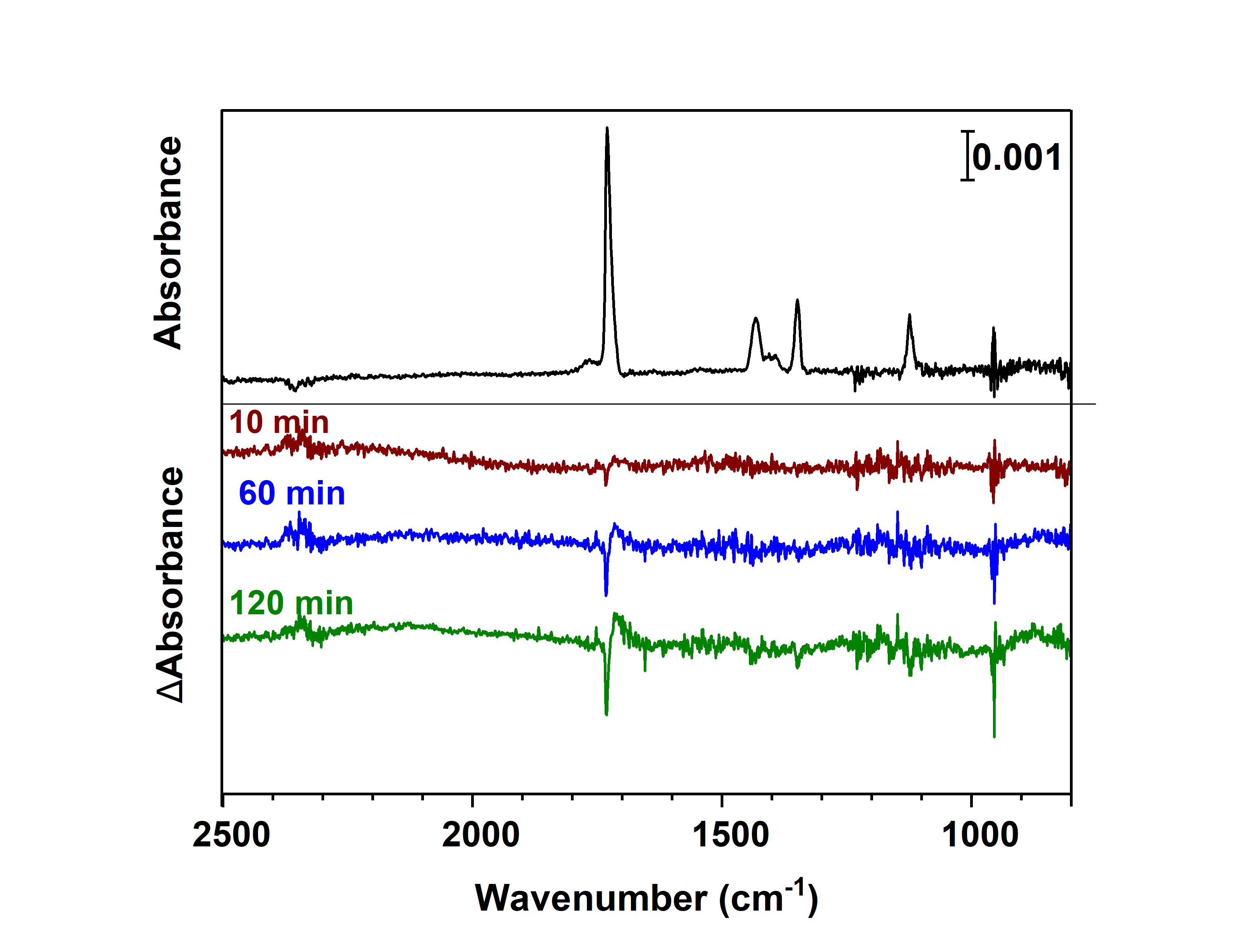}
   \caption{Variation in the difference spectra of the solid CH$_3$CHO on the Al substrate at 10~K after exposure to D atoms for 10, 60, and 120 min comparable with the initial CH$_3$CHO.}
   \label{fig:FTIR_CH3CHO+DonAlat10K}
\end{figure}

Figure \ref{fig:FTIR_CH3CHO+DonAlat10K} shows the evolution of the difference spectra of CH$_3$CHO after exposure to D atoms for up to 2 hours as well as that of CH$_3$CHO without atomic exposure. Similar to the behavior after exposure to H atoms, the C-O stretching band of CH$_3$CHO at 1728 cm$^{-1}$ decreased with exposure times, indicating the loss of CH$_3$CHO after interaction with D atoms on the Al substrate. Unfortunately, we cannot identify clear FTIR features for the products of the pre-deposited CH$_3$CHO with D atoms. 

\begin{figure}[ht!]
    \centering
    \includegraphics[width = 1.0\linewidth]{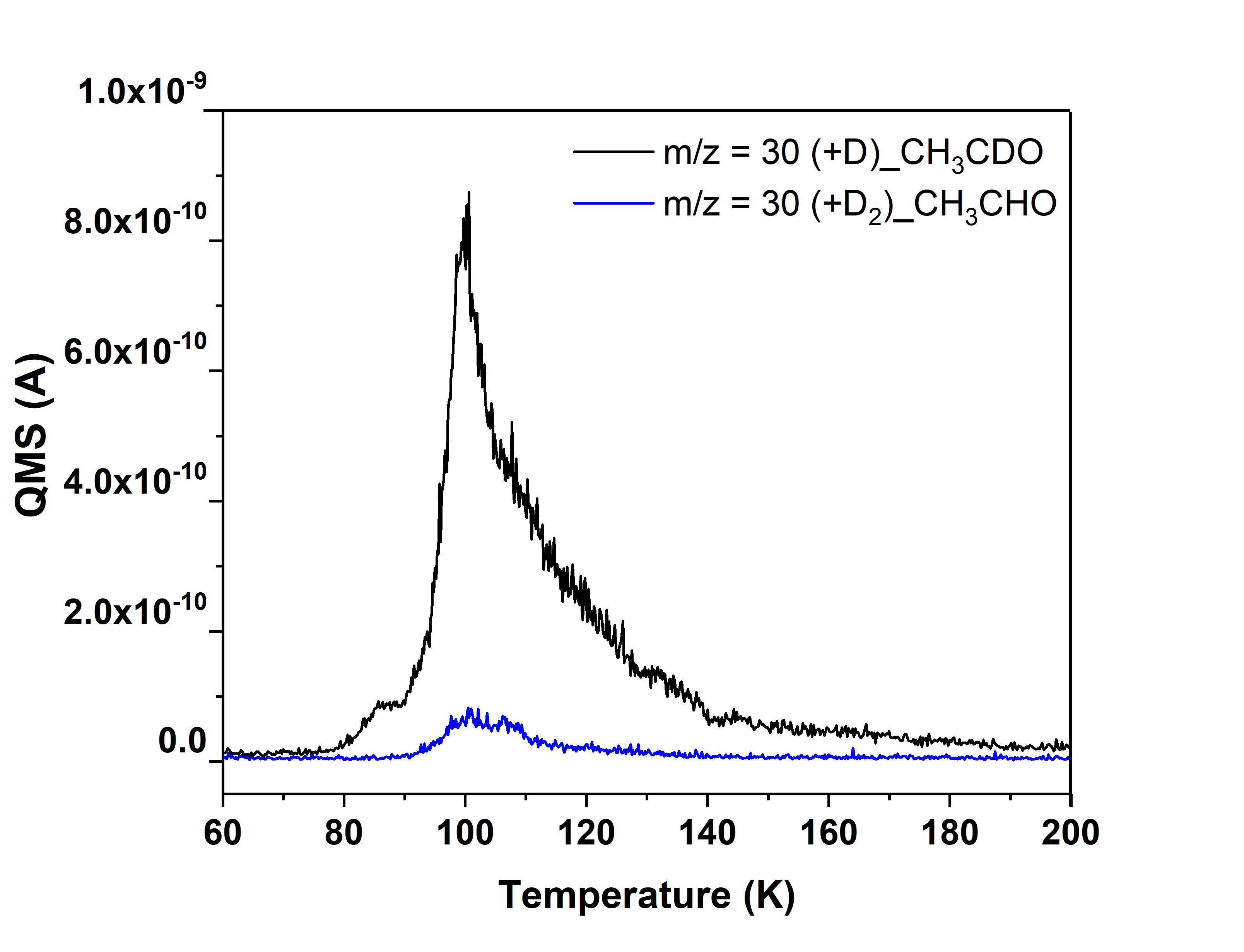}
    \includegraphics[width = 1.0\linewidth]{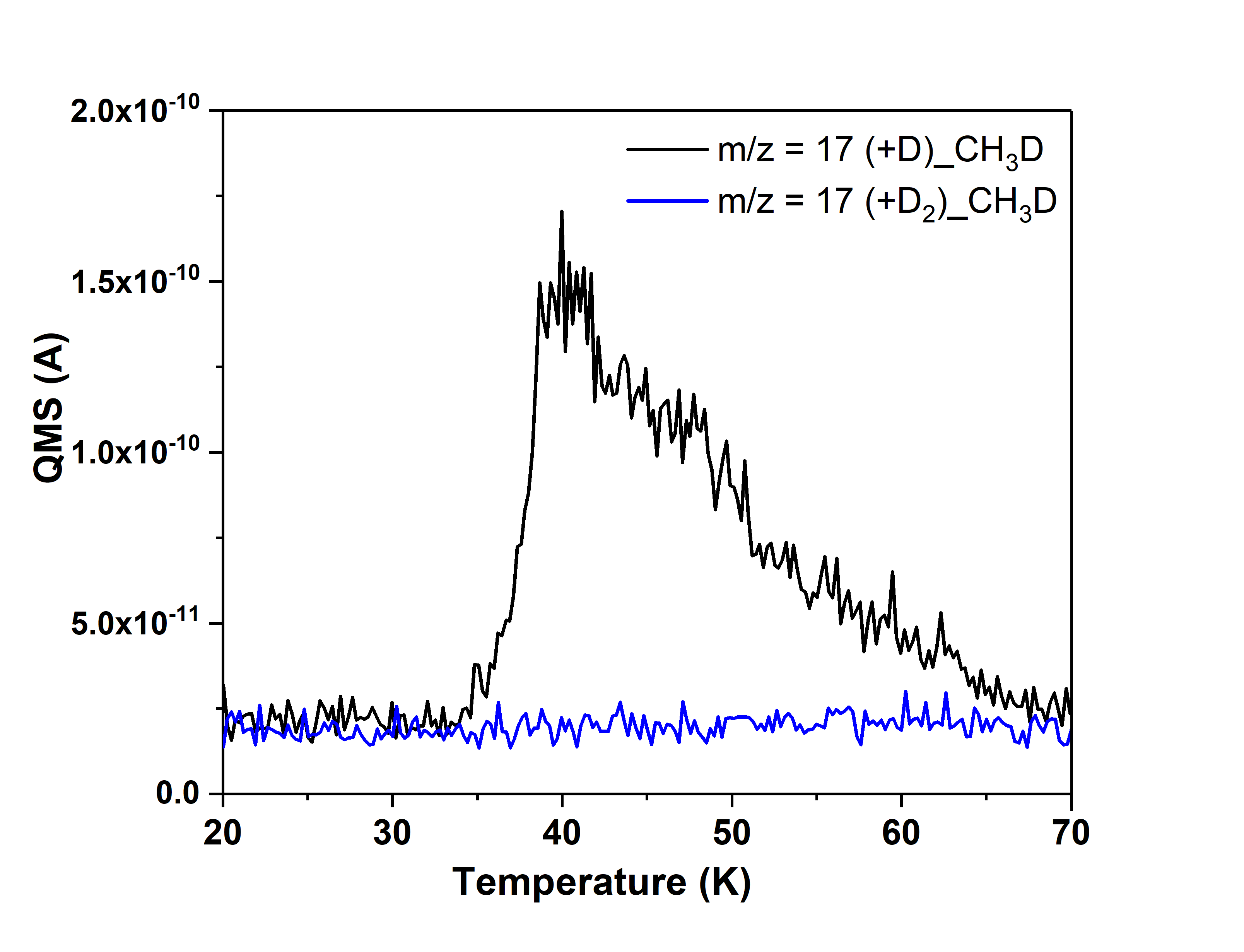}
    \includegraphics[width = 1.0\linewidth]{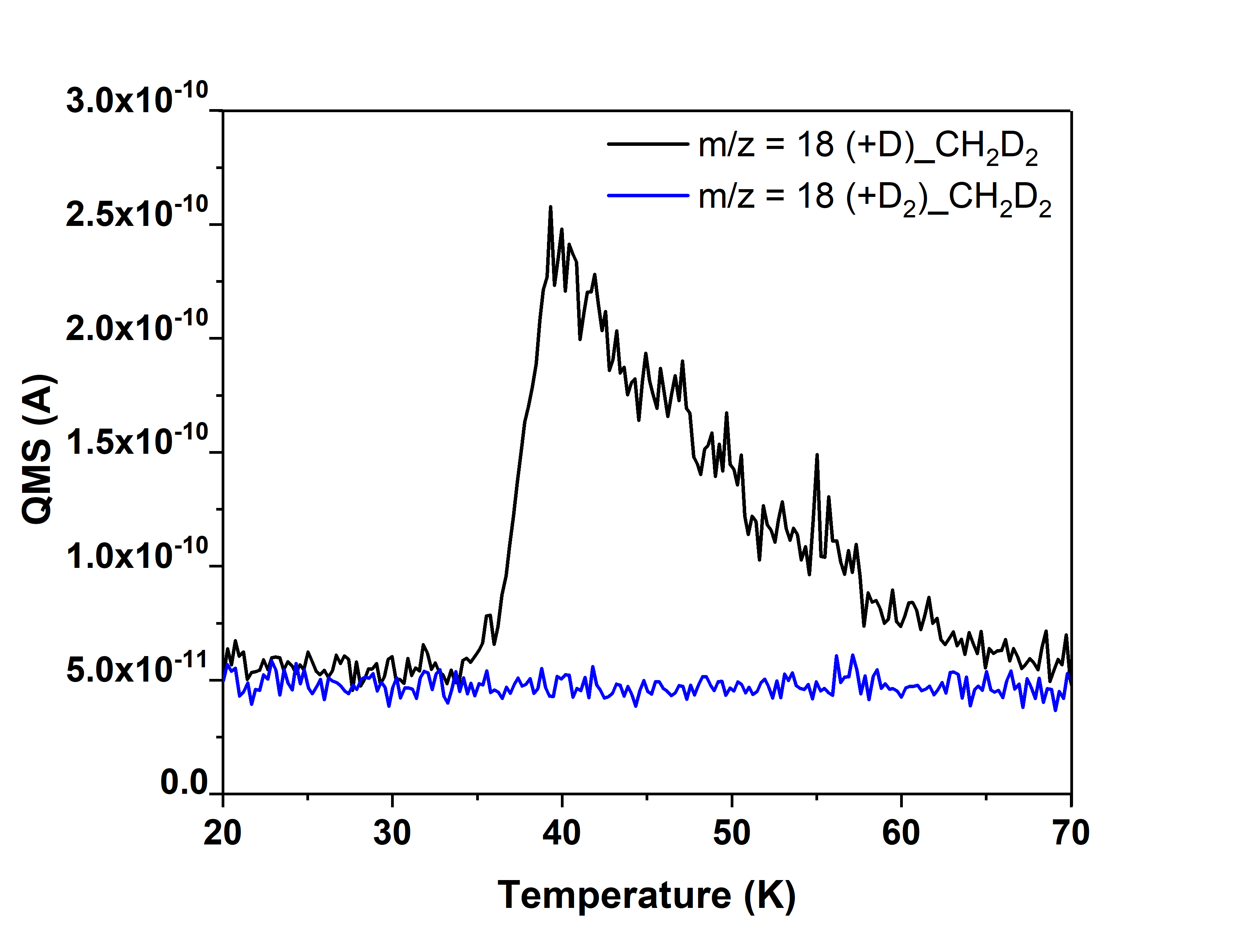}
    \caption{TPD-QMS profile of pre-deposition of CH$_3$CHO with D atoms (black line) comparable with D$_2$ (blue line) on Al substrate at 10~K, m/z = 30 (top panel); m/z = 17 (middle panel), and m/z = 18 (bottom panel). \rev{The m/z = 30 (top panel) in the blue line represents the fragmentation of \ce{CH3CHO}.}}
    \label{fig:TPD-QMS_CH3CHO+DonAl}
\end{figure}

On the other hand, we observed various features for the formation of isotopologues of the products based on the TPD experiments. Figure \ref{fig:TPD-QMS_CH3CHO+DonAl} shows the TPD spectrum after the reaction of CH$_3$CHO with D atoms on Al substrate at 10~K. Since CH$_3$CHO desorbs at the temperature between 90 - 145~K and shows a CHO fragment (m/z = 29) upon ionization, the desorption peak for m/z = 30 (the CDO fragment) is attributed to the deuterated counterpart of acetaldehyde, CH$_3$CDO. Thus, the desorption peak (m/z = 30 - top panel) observed at the same temperature after exposure to D atoms identified the formation of CH$_3$CDO. The formation of CH$_3$CDO provides compelling evidence for the reaction between CH$_3$CO and D atoms: CH$_3$CO~+D~$\xrightarrow{}$~CH$_3$CDO. This process, in which the CH$_3$CO radical is produced by the reaction \ref{eq:h1d}, is consistent with the calculated results.

Furthermore, the desorption peaks at m/z = 17 and 18 (Figure \ref{fig:TPD-QMS_CH3CHO+DonAl} - middle and bottom panels) within the temperature range of 35 - 65~K were indicative of the formation of deuterated methane isotopologues, CH$_3$D and CH$_2$D$_2$. These species could be produced by the reactions in \ref{eq:met} as discussed in section \ref{sec:merging}, although we cannot be certain due to the high barriers associated with H abstraction in \ce{CH3D} \citep{lamberts_methane_2022}. 

\begin{figure}[ht!]
       \centering
       \includegraphics[width = \linewidth]{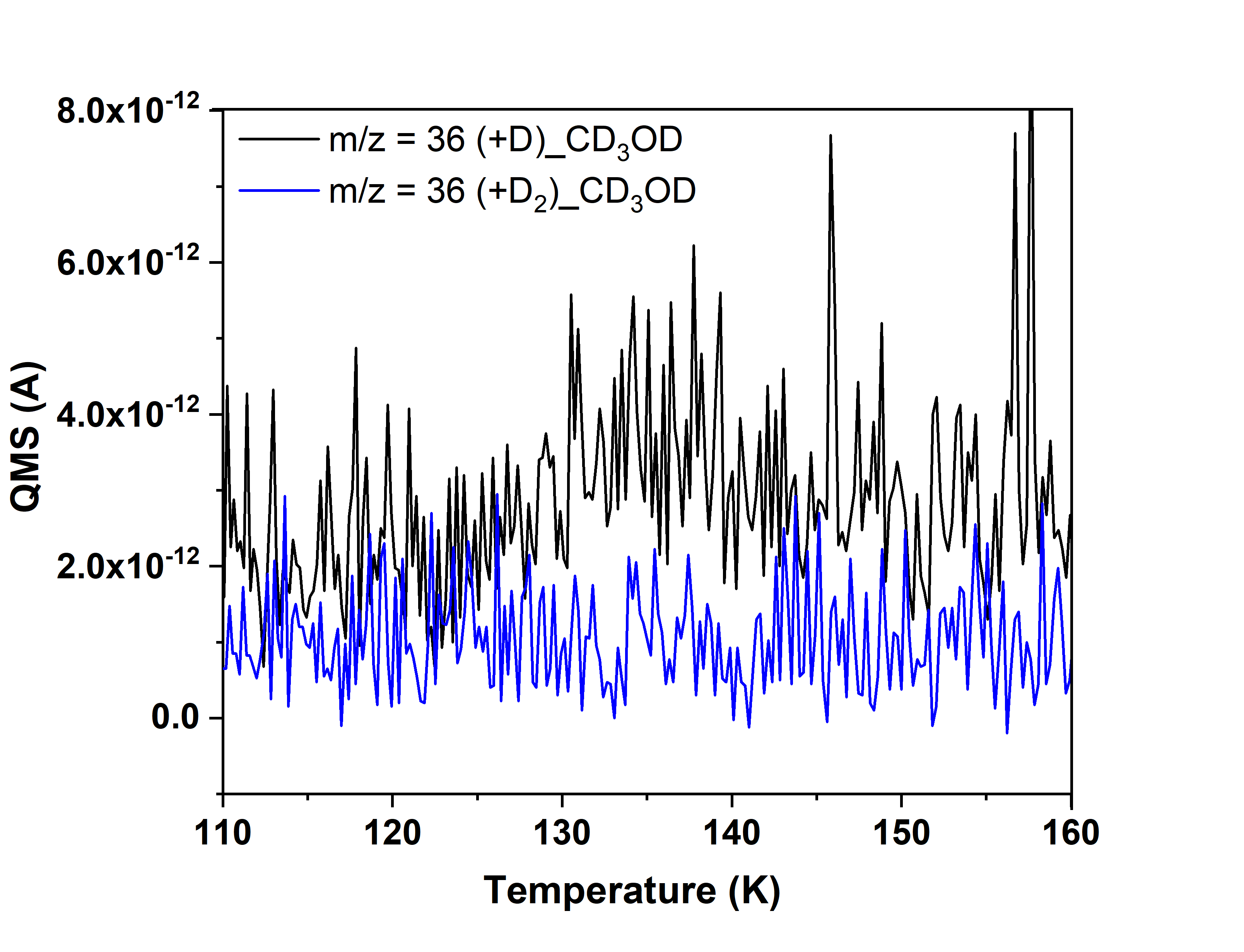}
       \caption{TPD-QMS profile of m/z = 36 observed for the pre-deposition of CH$_3$CHO and D atoms (black line) is compared with that of D$_2$ molecules (blue line) on the Al substrate.  }
       \label{fig:mass36ofCH3CHOwithDonAl}
   \end{figure}

\begin{figure}[ht!]
       \centering
       \includegraphics[width = \linewidth]{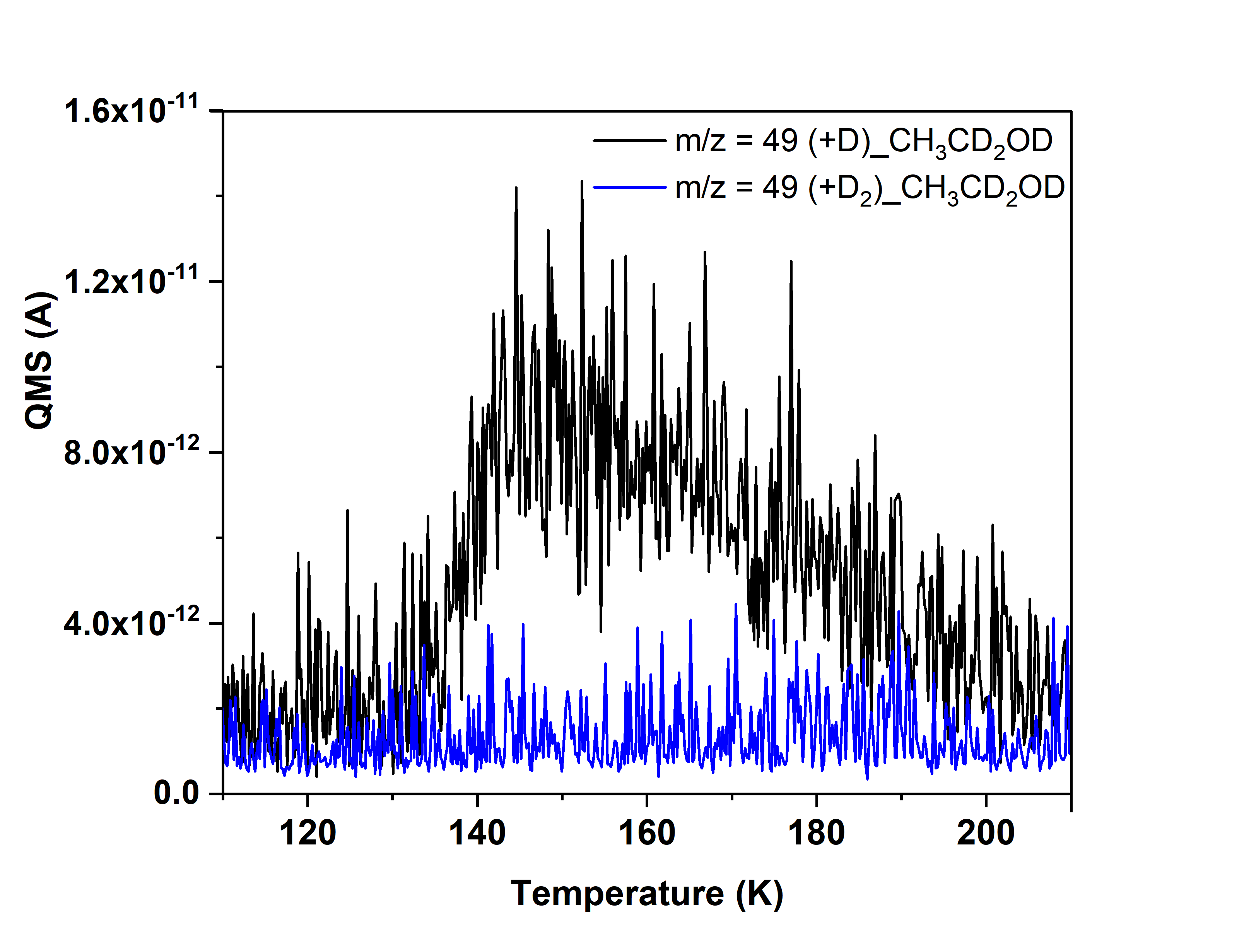}
       \includegraphics[width = \linewidth]{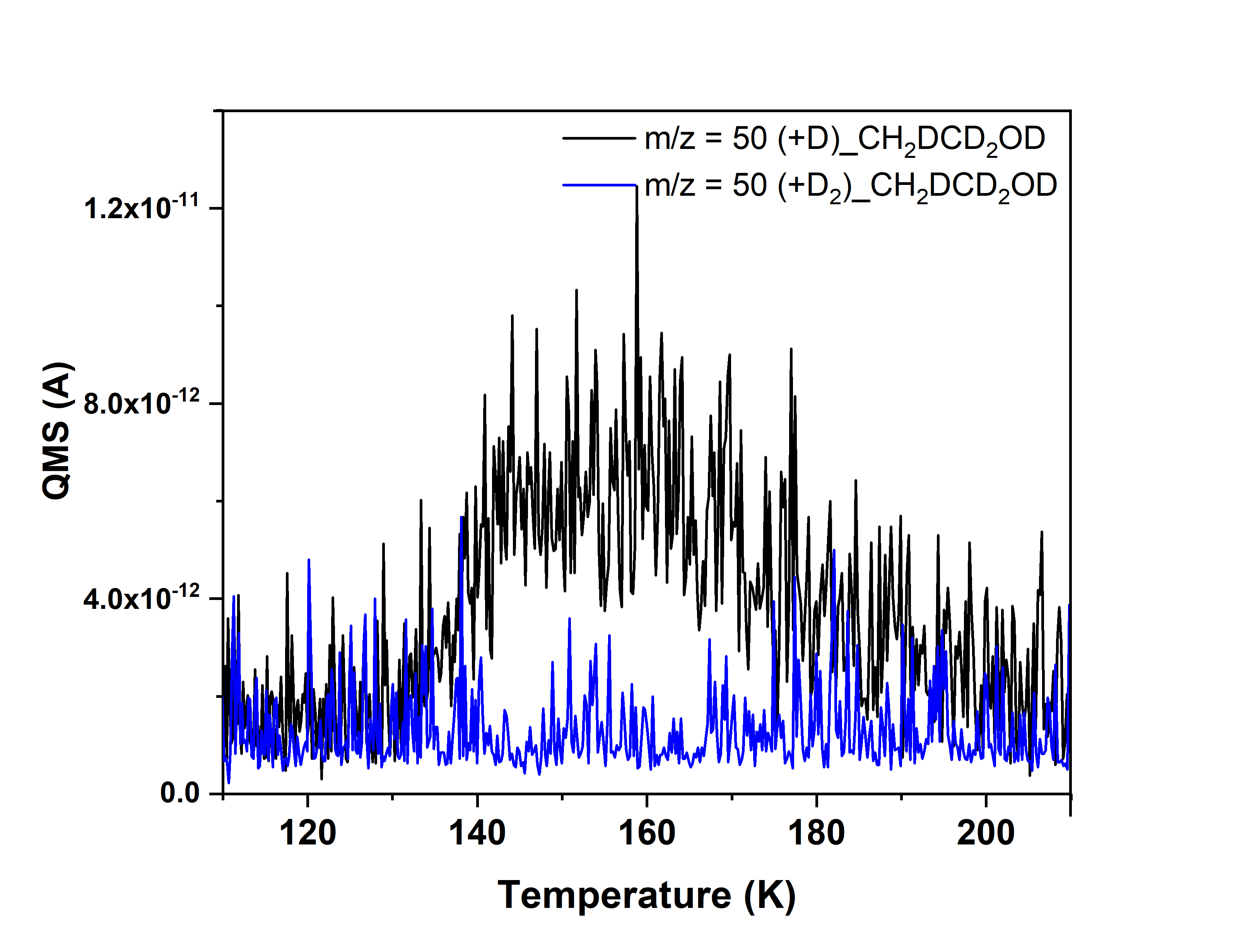}
       \caption{TPD-QMS profiles of m/z = 49 and 50 (black lines) desorbed at around 130 - 200 K, likely corresponding to ethanol isotopologues CH$_3$CD$_2$OD and/or CH$_2$DCD$_2$OD. These species  were yielded through the reaction of CH$_3$CHO and D atoms, with behavior comparable to that of D$_2$ molecules (blue line) on Al substrate at 10~K.}
       \label{fig:M49and50onAl}
   \end{figure}

A tiny peak corresponding to deuterated methanol (m/z = 36; CD$_3$OD) was observed to desorb at about 135~K (Figure \ref{fig:mass36ofCH3CHOwithDonAl}). The formation of CD$_3$OD has been proposed to result from the deuteration of CO \citep{Watanabe2002}, implying that the CO molecule is likely produced by the reaction of CH$_3$CO (reaction \ref{eq:ac-c5}) with H or D atoms under the current experimental conditions, a finding consistent with our calculated results. In addition, we also observed the desorption peaks at m/z = 49 and 50 in the temperature range of about 130 - 200~K. These peaks can be attributed to the formation of ethanol isotopologues corresponding to CH$_3$CD$_2$OD and/or CH$_2$DCD$_2$OD (see figure \ref{fig:M49and50onAl}), which were obtained by reactions of CH$_3$CO radicals with D atoms, followed by:

\begin{chequation} 
\begin{equation} \label{eq:thanh1}
   \rm{CH_3CO \xrightarrow[]{+D} H_2CCO \xrightarrow[]{+D} CH_2COD \xrightarrow{+3D} CH_2DCD_2OD}, 
   \end{equation}
\end{chequation}

\noindent and
\begin{chequation} 
\begin{equation} \label{eq:thanh2}
   \rm{CH_3CO \xrightarrow[]{+D} CH_3COD \xrightarrow[]{+D} CH_3CDOD \xrightarrow[]{+D} CH_3CD_2OD}.
\end{equation}   
\end{chequation}

The formation of ethanol isotopologues was consistent with the computational results obtained from reactions \ref{eq:ac-h3}  and \ref{eq:ac-o4} (see section \ref{sec:ch3o}) 
   
\begin{figure}[ht!]
       \centering
       \includegraphics[width = \linewidth]{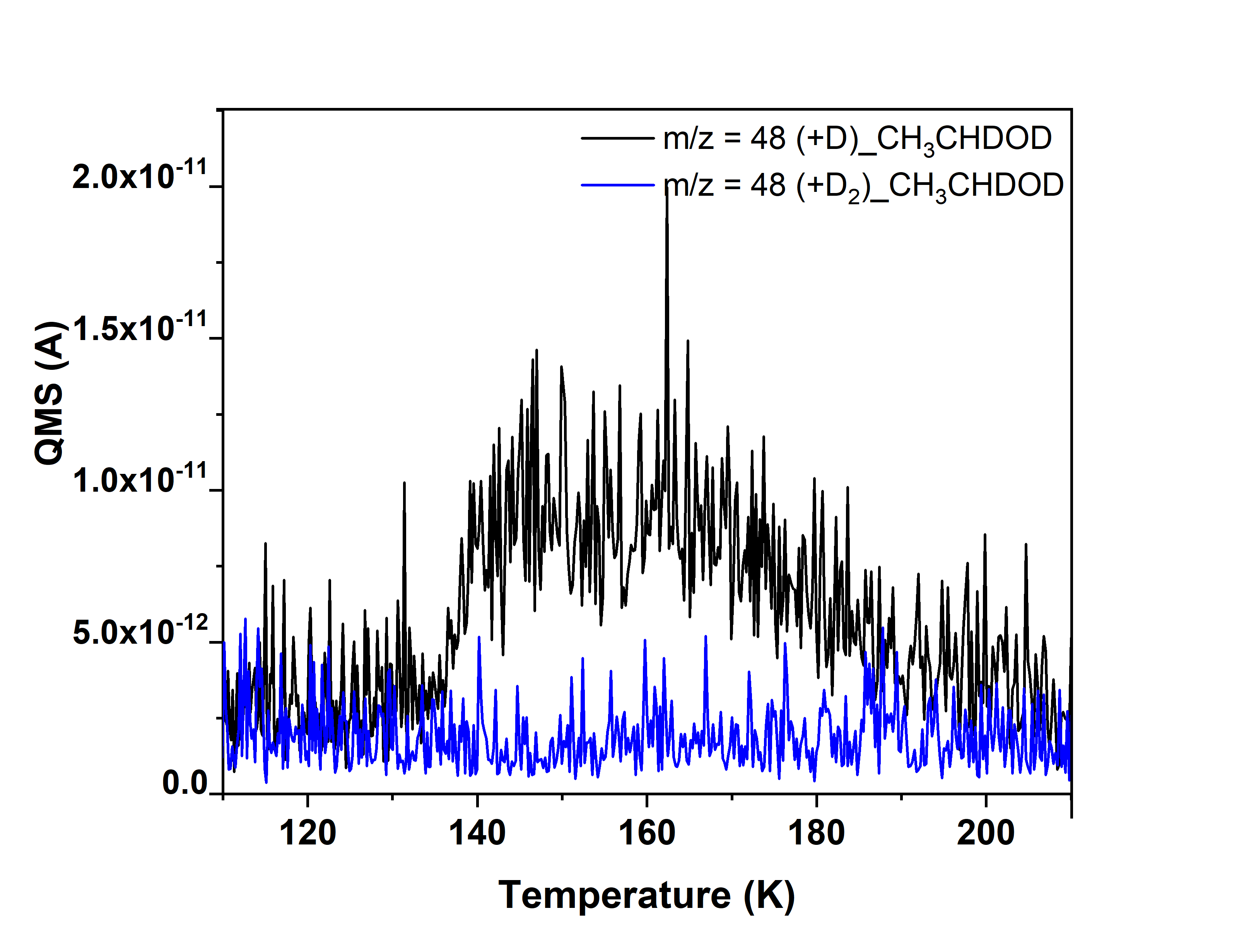}
       \caption{TPD-QMS profile for the desorption peak observed at m/z = 48 (black line) may be attributed to the formation of CH$_3$CHDOD through the addition reaction of D to CH$_3$CHO. This is in the comparison to the desorption behavior for D$_2$ (blue line) on the Al substrate at 10~K .}
       \label{fig:M48ofCH3CHOandD}
   \end{figure}

The appearance of a desorption peak observed at m/z = 48 (see Figure \ref{fig:M48ofCH3CHOandD}) likely indicated the formation of CH$_3$CHDOD, resulting from the addition reaction of D to CH$_3$CHO under the current experimental conditions. Nevertheless, based on the quantitative analysis of desorption areas for the reaction products, CH$_3$CHDOD was found to be a minor species in the reaction between CH$_3$CHO and D atoms. The predominant reaction involved CH$_3$CO radicals with D atoms, leading a higher yield of different products. 

\section{Discussion} \label{sec:discussion}

\subsection{Merging theory and experiments: Comparison with previous works} \label{sec:merging}

Once the theoretical and experimental results are discussed we are in a privileged position to discuss the chemical mechanism behind the surface hydrogenation of acetaldehyde. The main experimental finding comes after the analysis of the 1432, 1349, and 1124 cm$^{-1}$ IR decay bands that, after comparison with the amount of deposited \ce{CH3CHO}, reveals a limited destruction of around of a 10\%, with the rest of the \ce{CH3CHO} either unaffected or reformed after a H-abstraction and H-addition cycle. This finding is the easiest to rationalize based on the quantum chemical calculations.  The limited destruction/conversion of \ce{CH3CHO} is due to the dominance of the \ref{eq:h1} channel in the \ce{CH3CHO + H/D} reaction. Previous literature studies on the hydrogenation of aldehydes \citep{mondal_is_2021} have not considered the H-abstraction reactions, perhaps in analogy to the limited importance of this type of reaction in the parent molecule formaldehyde, \ce{H2CO}. \citep{song_tunneling_2017}. However, for \ce{CH3CHO} we found otherwise, with H-abstraction dominating the reactivity. Our deuteration experiments showed a tiny fraction of \ce{CH$_3$CHDOD} indicating also the possibility of some hydrogen/deuterium addition in the hydrogenation/deuteration of \ce{CH3CHO}. This is somewhat coherent with the rate constants for H addition found in reactions \ref{eq:c2} and \ref{eq:c2d}, when \ce{CH3CHO} and \ce{H} are in deep binding sites with slow diffusion rates (See Figure \ref{fig:theo5} and associated discussion). However, H-addition is a minor channel of the reaction, whose quantification to the total branching ratio is difficult. Therefore, in Section \ref{sec:recommendations} we recommend considering a 1.0 branching ratio toward H-abstraction (Reaction \ref{eq:h1}). 

After reaction \ref{eq:h1}, any subsequent hydrogenation on the new system involves the very reactive \ce{CH3CO} radical. From a theoretical standpoint, we determined that all reaction positions are, in principle, possible. Beginning with reaction \ref{eq:ac-c2}, the H (or D) addition on this position reforms \ce{CH3CHO}, and based on the experiment, is the most likely outcome, with an 90 \% of the reactive events in this direction, determined from the total fraction of \ce{CH3CHO} remaining after concluding our experiments. However, neither our calculations nor our experiments can rule out the contribution of the reforming of \ce{CH3CHO} from secondary reactions, i.e. hydrogenation of \ce{H2CCO} \cite[see below and][]{ibrahim_formation_2022,ferrero_formation_2023,ibrahim_significant_2024}. For reactions \ref{eq:ac-h3}, \ref{eq:ac-o4} and \ref{eq:ac-c5} the products are \ce{H2CCO} (and \ce{H2}), \ce{CH3COH} (methyl-methoxy carbene; \cite{schreiner_methylhydroxycarbene_2011}), and \ce{CH4 + CO}. Beginning with \ce{H2CCO}, as we mention in Section \ref{sec:experiments}, we cannot identify it in the IR spectrum because its main transition falls at 2133 cm$^{-1}$ overlapping with the CO band. However, the appearance of additional desorption features for m/z = 14 (\ce{CH2}) and 42 (\ce{H2CCO}) between 75--90 K and 150--165 K in the TPD experiments (Figure \ref{fig:keteneevidence}) is in line with previous studies \citep{maity_formation_2015,Chuang2020,Fedoseev2022}, which confirmed the desorption of \ce{H2CCO}. Nevertheless, \citet{ferrero_formation_2023}, using instanton calculations like the ones shown in this work, showed that the \ce{H2CCO} hydrogenation is very efficient with final products \ce{CH3CHO} and perhaps \ce{C2H5OH} (ethanol). \citet{Bisschop2007,ibrahim_formation_2022} also report \ce{CO + CH4} and \ce{H2CO} as products of the \ce{H2CCO + H} reaction, which is also coherent with our results. Combining experiments and theory, we determined that \ce{H2CCO} must be a product of the \ce{CH3CHO} hydrogenation under interstellar conditions, but in small amounts. Continuing with \ce{CH3COH}, even though it is formed in our calculations, we are unable to comment much on this species from an experimental point of view. This is due to the limited experimental data, but also to its transient nature. As a reactive carbene, we expect that the \ce{CH3COH + 2H -> C2H5OH} reaction will dominate in our experimental setup. Indeed, \ce{C2H5OH} is seen in our co-deposition experiments, and the experimental and theoretical synergy carried out in this work allowed us that it is a secondary product of the hydrogenation of \ce{CH3CHO} rather than a direct one. The presence of \ce{C2H5OH} was puzzling when checking the previous literature. In the experiments of \citet{Fedoseev2022} (for \ce{H2CCO + H}) \ce{C2H5OH} was seen, as in our experiments. However, \citet{ibrahim_formation_2022} and \citet{ibrahim_significant_2024} do not report its presence after the same reaction. The reason for this disagreement is likely due to the experimental conditions, and our results provide a reconciling scenario. We could not detect \ce{C2H5OH} using the IR spectra in our pre-deposition experiment, which would be the one directly comparable to the experiments of \citet{ibrahim_formation_2022} needing to rely on TPD-QMS measurements (See Figure \ref{fig:TPD-QMS_for_predeposition}) for that. However, the presence of \ce{C2H5OH} was seen in our co-deposition experiments starting from \ce{CH3CHO} mimicking \citet{Fedoseev2022}. Although other experimental conditions differ between our setups, like H flux or irradiation time, we think the reason for the disagreement stems from the deposition strategy.

Finally, the last experimental product that we observe is \ce{CH4} (and its isotopologues, see below) and CO. Given the enormous $\Delta H^{ \ddagger}$ found for reaction \ref{eq:c5}, the formation of \ce{CH4} must come from the hydrogenation of \ce{CH3CO}, or perhaps of further radicals in the hydrogenation sequence of \ce{CH3CO}, out of the scope of the present paper. However, we find it likely that \ce{CH4} comes directly from \ce{CH3CO} via reaction \ref{eq:ac-c5}. CO was seen in our experiments, either directly as CO or as its hydrogenated products \ce{H2CO} and \ce{CH3OH} as discussed above.

Our deuteration experiments do show a variety of isotopologues of the products discussed in the previous paragraphs. The presence or absence of certain isotopologues not only aids our experiments but also allows us to confirm our theoretical observations. For \ce{CH3CHO + D} our experiments reveal the majoritarian formation of \ce{CH3CO} that later reforms \ce{CH3CDO}. Likewise, we clearly see the formation of \ce{CH3D} and \ce{CH2D2}, that might arise from the sequence of reactions (omitting \ce{HD} when applicable):

\begin{chequation}                          
   \begin{equation} \label{eq:met}
   \resizebox{\columnwidth}{!}{
   \ce{CH3CHO ->[D] CH3CO ->[D] CH3D + CO ->[D] CH2D ->[D] CH2D2}}
   \end{equation}      
\end{chequation}

These reactions serve as a proxy to suggest that the \ce{CH3D + H/D -> CH2D + H2 / HD} could be a viable reaction, although we cannot confidently ensure it, because some reaction steps in \ref{eq:met} have high barriers. A preliminary search at the rev-DSD-PBEP86(D4)/jun-cc-pV(T+d)Z level for the \ce{CH3D + D -> CH2D + HD}~(The third reaction in the scheme of reactions shown in \ref{eq:met}) shows that the reactions is slightly exothermic, unlike the full hydrogenated counterpart, by $\Delta H^{ \textrm{R}}$=-3.3 kJ mol$^{-1}$ (-404 K), making the reaction scheme \ref{eq:met} in principle viable albeit probably very slow. Overall, while the presence of \ce{CH3D} is easy to explain from experiments and theory, the presence of \ce{CH2D2} is not that easily explained, with secondary and tertiary hydrogenation branches being important, and requiring a careful validation out of the scope of this work. The CO formed in the previous reaction can also be hydrogenated to fully deuterated methanol (\ce{CD3OD}) \citep{Watanabe2002}, a reaction that could not happen without the formation of CO, further increasing our confidence in the viability of reaction \ref{eq:ac-c5}. We also observe clear signals of the formation of ethanol isotopologues, especially of \ce{CH3CD2OD} or \ce{CH2DCD2OD} (Reaction scheme in \ref{eq:thanh1} and \ref{eq:thanh2} and Figure \ref{fig:M49and50onAl}), merging once again experimental and theoretical results and highlighting the importance of reaction \ref{eq:h1} as the initiator of the complex chemistry observed in the experiments. We reiterate that a small fraction of \ce{CH3CHDOD} was also found (Figure \ref{fig:M48ofCH3CHOandD}).

We briefly comment on the differences between our new set of experiments and the classical ones of \citet{Bisschop2007}, that inspired this work. In both our experiments we obtain the same products of the reaction, although in different proportions. While we only find a conversion of \ce{CH3CHO} equal to 10\% in the duration of our experiment, \citet{Bisschop2007} reports variable conversion rates varying between 30\% to almost total conversion (See their Table 5). These quantitative differences should come from different factors, being the most likely reasons the temperature of the H atoms and the selection of substrate. In our setup, the H atoms are cooled down before hitting the surface, guaranteeing immediate thermalization on the surface. In \citet{Bisschop2007} the H-atoms were introduced in the chamber at 300 K, which can promote a plethora of reactions through different reaction mechanisms like the Eley-Rideal or Harris-Kasemo ones. Moreover, in our setup, we carry our experiments on ASW where the interactions between \ce{CH3CHO} molecules with the surface will be different than in pure \ce{CH3CHO} or in a metallic substrate.

Lastly, in light of what it has been presented in this section we have prepared a series of recommendations to include this reaction in astrochemical rate equation models. We have gathered our conclusions in Appendix \ref{sec:recommendations}.

\subsection{Astrophysical Implications} \label{sec:implications}

Our results carry a series of implications for the astrochemical community, in particular regarding the deuteration and reactive desorption of \ce{CH3CHO}. Acetaldehyde is a molecule routinely detected in a variety of interstellar environments \citep{Cazaux2003, occhiogrosso_ethylene_2014, walsh_complex_2014, Imai2016, Codella2015, Holdship2019, bonfand_complex_2019}, including several detections in prestellar cores \citep{bacmann_detection_2012, cernicharo_discovery_2012, Scibelli2021, Jimenez_Serra_2016,megias_complex_2022}, where there is little energy for bringing the molecules to the gas phase, where they are detected. The formation of \ce{CH3CHO} is debated, including the gas-phase route \ce{C2H5 + O -> CH3CHO + H} \citep{vazart_gas-phase_2020} or ice surface routes. These routes involve the radical-radical recombination of \ce{HCO + CH3}, on ice \citep{Garrod2013,Molpeceres2024com,Lamberts2019} although on water ice this route is sometimes considered less effective \citep{enrique-romero_theoretical_2021}. Alternative routes include the radiolytic \citep{shingledecker_cosmic-ray-driven_2018} or non-thermal formation \citep{Jin2020}. Regardless of the formation route, \ce{CH3CHO} will form or deplete on the grain, and it is at this stage when the title reaction can take place.

Our results show that acetaldehyde is mostly processed and reformed through reactions \ref{eq:h1} and \ref{eq:ac-c2}, and although other products and routes exist, as extensively discussed in this manuscript, the astrophysical implications of our work are mainly carried by this limited destruction, and we will focus on those in this section. In the first place, we talk about reactive desorption. When an H atom is abstracted from the HCO moiety in \ce{CH3CHO}, the hydrogenation of the resulting \ce{CH3CO} radical could trigger a chemical desorption event, considering that the binding energy of \ce{CH3CHO} is among the lowest for COMs. As we indicated in Section \ref{sec:experiments} we cannot confirm this possibility, as the different hydrogenation branches opened from our reaction make it very difficult to confirm or deny the possibility of chemical desorption. Optimistic estimates of the probability of chemical desorption for COMs place the probability per reaction event at around 1\% \citep{garrod_non-thermal_2007}. Because this amount should be included in the 10\% of nonreformed \ce{CH3CHO} it is experimentally extremely hard to prove. Yet, \ce{CH3CHO} constitutes the best-case scenario with an effective H-abstraction / H-Addition cycle, the high reaction energy for the \ce{CH3CO + H} reaction and the relatively low binding energy on water ice \citep{molpeceres_desorption_2022,ferrero_acetaldehyde_2022}.

Another important implication of our work pertains to the formation of ethanol, \ce{C2H5OH}, which is predicted to be formed in very low amounts from reaction \ref{eq:c2}, but actually from the evolution of the \ce{CH3CO + H} reaction. This is in contrast with the hydrogenation of formaldehyde (\ce{H2CO}), the prototypical interstellar aldehyde, that leads to methanol \ce{CH3OH} \citep{Watanabe2002,Hidaka2004}. In \citet{mondal_is_2021} the authors indicate that the connection from \ce{CH3CHO} with \ce{C2H5OH} through direct hydrogenation is not enough to account for the abundances of \ce{C2H5OH} in the hot core G10.47+0.03, and that their models revealed that \ce{CH3 + CH2OH -> C2H5OH} is needed to explain the observations. Our results fully adhere to this picture and strongly suggest that the link between \ce{CH3CHO} and \ce{C2H5OH} is not straightforward.

Finally, we reserve some words for the possibility of finding deuterated acetaldehyde. A relatively recent publication \citet{jorgensen_alma-pils_2018} showed that \ce{CH3CDO} is the isotopologue of a COM with the highest ratio to the non-deuterated molecule, with a percentage of 8\% in IRAS16293–2422. This result reinforces the results obtained in this work and suggests as a possible explanation for this observation the H-abstraction / H-addition route discussed in this work, in addition to the explicit detection of \ce{CH3CDO} in our experiments. Nevertheless, it is important to remark that other \ce{CH3CHO} isotopologues \ce{CH2DCHO} and \ce{CD2HCHO} have also been detected in the same source \citep{coudert_astrophysical_2019, ferrer_asensio_millimetre_2023}. These last findings do not rule out the importance of our reaction routes but show the complexity associated with deuterium fractionation in the ISM, where gas and surface chemistry are in constant interplay. 

\section{Conclusions} \label{sec:conclusion}

We conclude by providing a series of bullet points that summarize our work.

\begin{enumerate}
   \item Our calculations and experiments show that the \ce{CH3CHO + H} does not lead to significant conversion. The reason for that limited conversion is found in the dominance of hydrogen abstraction in the HCO moiety of \ce{CH3CHO} which in turn leads to facile reforming of \ce{CH3CHO}. Quantifying this conversion leads us to a maximum of 10\% of \ce{CH3CHO} under our experimental conditions.
   \item Other minor products of the reaction of \ce{CH3CHO} with H come from the hydrogenation of the \ce{CH3CO} reactive radical and are \ce{H2CO}, \ce{CH3OH}, \ce{C2H5OH}, \ce{H2CCO}, CO or \ce{CH4} revealing a very complex chemistry. In general, our calculations and previous knowledge of the chemistry of the different intermediates found in our calculations can satisfactorily explain the different products that we obtain. However, and because the study of the \ce{CH3CO + H} is more qualitative due to theoretical constraints, more sophisticated approaches are required to derive accurate branching ratios for the reaction.
   \item Our experiments reveal the effective formation of \ce{CH3CDO} supporting the idea of a H/D-abstraction and H/D-addition cycle as the major route for the \ce{CH3CHO + H} reaction.
   \item Although we can explain the vast majority of our experimental observations with our calculations, the presence of doubly deuterated methane \ce{CH2D2} and the \ce{CH3CHDOD} isotopologue in our deuteration experiments remains difficult to explain. In general, several hydrogenation and deuteration branches of the minor products could lead to different products making explaining the complete reaction network an unbearable task.
   \item Our results align with the results found in the modeling literature \citep{mondal_is_2021} suggesting that the link between acetaldehyde and ethanol is not direct from the former to the latter via hydrogenation.
   \item We were unable to determine whether reactive desorption happens for \ce{CH3CHO} although in light of the energetic parameters found by our theoretical calculations, summed to the H-abstraction and addition cycle present in the reaction, we consider it a likely outcome.
   \item Our experiments and calculations nicely explain the excess of \ce{CH3CDO} found in the IRAS16293–2422 hot core \citep{jorgensen_alma-pils_2018}. However, we remark that a single chemical reaction is not enough to explain complex deuterium enrichment processes in space.
\end{enumerate}

Future avenues of the present work could address the presence of certain non-easy-to-explain isotopologues in our experiments, or the study of the impact of the here-derived quantities in realistic astrochemical models. Likewise, we are currently tackling the complex problem of \ce{CH3CHO} chemical desorption using more sophisticated molecular dynamics approaches. We will continue combining experiments and calculations to fully unravel complex interstellar reaction networks.

\begin{acknowledgements}
   \rev{The authors thank the anonymous reviewer for the thorough revision of the manuscript.} We thank Professor Johannes K\"astner at Stuttgart University for providing a development version of \textsc{ChemShell}. We also thank Dr. Masashi Tsuge and Dr. Hiroshi Hidaka for fruitful discussions on the experimental results. G.M acknowledges the support of the grant RYC2022-035442-I funded by MICIU/AEI/10.13039/501100011033 and ESF+. G.M. also received support from project 20245AT016 (Proyectos Intramurales CSIC). We acknowledge the computational resources provided by bwHPC and the German Research Foundation (DFG) through grant no INST 40/575-1 FUGG (JUSTUS 2 cluster), the DRAGO computer cluster managed by SGAI-CSIC, and the Galician Supercomputing Center (CESGA). The supercomputer FinisTerrae III and its permanent data storage system have been funded by the Spanish Ministry of Science and Innovation, the Galician Government and the European Regional Development Fund (ERDF). Y.O and N.W. acknowledge the funding support from JSPS KAKENHI grant nos. JP23H03980, JP21H04501, and JP22H00159.  
\end{acknowledgements}

%
%

\bibliographystyle{aa}
\bibliography{bib.bib}

\begin{thebibliography}{97}
\expandafter\ifx\csname natexlab\endcsname\relax\def\natexlab#1{#1}\fi

\bibitem[{Alvarez-Barcia {et~al.}(2018)Alvarez-Barcia, Russ, Kästner, \& Lamberts}]{Alvarez-Barcia2018}
Alvarez-Barcia, S., Russ, P., Kästner, J., \& Lamberts, T. 2018, Monthly Notices of the Royal Astronomical Society, 479, 2007, arXiv: 1806.02062

\bibitem[{Asgeirsson {et~al.}(2017)Asgeirsson, Jónsson, \& Wikfeldt}]{asg17}
Asgeirsson, V., Jónsson, H., \& Wikfeldt, K.~T. 2017, Journal of Physical Chemistry C, 121, 1648

\bibitem[{{Bacmann} \& {Faure}(2014)}]{Bacmann2014}
{Bacmann}, A. \& {Faure}, A. 2014, in SF2A-2014: Proceedings of the Annual meeting of the French Society of Astronomy and Astrophysics, ed. J.~{Ballet}, F.~{Martins}, F.~{Bournaud}, R.~{Monier}, \& C.~{Reyl{\'e}}, 3--8

\bibitem[{Bacmann {et~al.}(2012)Bacmann, Taquet, Faure, Kahane, \& Ceccarelli}]{bacmann_detection_2012}
Bacmann, A., Taquet, V., Faure, A., Kahane, C., \& Ceccarelli, C. 2012, Astronomy \& Astrophysics, 541, L12

\bibitem[{Barone \& Cossi(1998)}]{barone_quantum_1998}
Barone, V. \& Cossi, M. 1998, The Journal of Physical Chemistry A, 102, 1995

\bibitem[{Bartlett \& Purvis(1978)}]{bartlett_manybody_1978}
Bartlett, R.~J. \& Purvis, G.~D. 1978, International Journal of Quantum Chemistry, 14, 561

\bibitem[{{Bennett} {et~al.}(2005){Bennett}, {Jamieson}, {Osamura}, \& {Kaiser}}]{Bennet2005}
{Bennett}, C.~J., {Jamieson}, C.~S., {Osamura}, Y., \& {Kaiser}, R.~I. 2005, \apj, 624, 1097

\bibitem[{{Bisschop} {et~al.}(2007){Bisschop}, {Fuchs}, {van Dishoeck}, \& {Linnartz}}]{Bisschop2007}
{Bisschop}, S.~E., {Fuchs}, G.~W., {van Dishoeck}, E.~F., \& {Linnartz}, H. 2007, Astron. Astrophys., 474, 1061

\bibitem[{Bohlin {et~al.}(1978)Bohlin, Savage, \& Drake}]{bohlin_survey_1978}
Bohlin, R.~C., Savage, B.~D., \& Drake, J.~F. 1978, The Astrophysical Journal, 224, 132

\bibitem[{Bonfand {et~al.}(2019)Bonfand, Belloche, Garrod, Menten, Willis, Stéphan, \& Müller}]{bonfand_complex_2019}
Bonfand, M., Belloche, A., Garrod, R.~T., {et~al.} 2019, Astronomy \& Astrophysics, 628, A27

\bibitem[{Caldeweyher {et~al.}(2019)Caldeweyher, Ehlert, Hansen, Neugebauer, Spicher, Bannwarth, \& Grimme}]{Caldeweyher2019}
Caldeweyher, E., Ehlert, S., Hansen, A., {et~al.} 2019, Journal of Chemical Physics, 150

\bibitem[{Cazaux {et~al.}(2003)Cazaux, Tielens, Ceccarelli, Castets, Wakelam, Caux, Parise, \& Teyssier}]{Cazaux2003}
Cazaux, S., Tielens, A. G. G.~M., Ceccarelli, C., {et~al.} 2003, Astrophys. J., 593, L51

\bibitem[{{Ceccarelli} {et~al.}(2014){Ceccarelli}, {Caselli}, {Bockel{\'e}e-Morvan}, {Mousis}, {Pizzarello}, {Robert}, \& {Semenov}}]{Ceccareli2014}
{Ceccarelli}, C., {Caselli}, P., {Bockel{\'e}e-Morvan}, D., {et~al.} 2014, in Protostars and Planets VI, ed. H.~{Beuther}, R.~S. {Klessen}, C.~P. {Dullemond}, \& T.~{Henning}, 859--882

\bibitem[{Cernicharo {et~al.}(2012)Cernicharo, Marcelino, Roueff, Gerin, Jiménez-Escobar, \& Muñoz~Caro}]{cernicharo_discovery_2012}
Cernicharo, J., Marcelino, N., Roueff, E., {et~al.} 2012, The Astrophysical Journal, 759, L43

\bibitem[{Chang {et~al.}(2007)Chang, Cuppen, \& Herbst}]{changGasgrainChemistryCold2007}
Chang, Q., Cuppen, H.~M., \& Herbst, E. 2007, Astronomy \& Astrophysics, 469, 973

\bibitem[{Chuang {et~al.}(2020)Chuang, Fedoseev, Ioppolo, van Dishoeck, \& Linnartz}]{Chuang2020}
Chuang, K.~J., Fedoseev, G., Ioppolo, S., van Dishoeck, E.~F., \& Linnartz, H. 2020, Monthly Notices of the Royal Astronomical Society: Letters, 455, 1702

\bibitem[{Codella {et~al.}(2015)Codella, Fontani, Ceccarelli, Podio, Viti, Bachiller, Benedettini, \& Lefloch}]{Codella2015}
Codella, C., Fontani, F., Ceccarelli, C., {et~al.} 2015, Mon. Not. R. Astron. Soc. Lett., 449, L11

\bibitem[{Coudert {et~al.}(2019)Coudert, Margulès, Vastel, Motiyenko, Caux, \& Guillemin}]{coudert_astrophysical_2019}
Coudert, L.~H., Margulès, L., Vastel, C., {et~al.} 2019, Astronomy \& Astrophysics, 624, A70

\bibitem[{Dunning(1989)}]{Dunning1989}
Dunning, T.~H. 1989, The Journal of Chemical Physics, 90

\bibitem[{Eckart(1930)}]{Eckart1930}
Eckart, C. 1930, Physical Review, 35, 1303, publisher: American Physical Society ISBN: 0031-899X

\bibitem[{Enrique-Romero {et~al.}(2021)Enrique-Romero, Ceccarelli, Rimola, Skouteris, Balucani, \& Ugliengo}]{enrique-romero_theoretical_2021}
Enrique-Romero, J., Ceccarelli, C., Rimola, A., {et~al.} 2021, Astronomy \& Astrophysics, 655, A9

\bibitem[{Enrique-Romero {et~al.}(2022)Enrique-Romero, Rimola, Ceccarelli, Ugliengo, Balucani, \& Skouteris}]{enrique-romero_quantum_2022}
Enrique-Romero, J., Rimola, A., Ceccarelli, C., {et~al.} 2022, The Astrophysical Journal Supplement Series, 259, 39

\bibitem[{{Fedoseev} {et~al.}(2022){Fedoseev}, {Qasim}, {Chuang}, {Ioppolo}, {Lamberts}, {van Dishoeck}, \& {Linnartz}}]{Fedoseev2022}
{Fedoseev}, G., {Qasim}, D., {Chuang}, K.-J., {et~al.} 2022, \apj, 924, 110

\bibitem[{Ferrer~Asensio {et~al.}(2023)Ferrer~Asensio, Spezzano, Coudert, Lattanzi, Endres, Jørgensen, \& Caselli}]{ferrer_asensio_millimetre_2023}
Ferrer~Asensio, J., Spezzano, S., Coudert, L.~H., {et~al.} 2023, Astronomy \& Astrophysics, 670, A177

\bibitem[{Ferrero {et~al.}(2023)Ferrero, Ceccarelli, Ugliengo, Sodupe, \& Rimola}]{ferrero_formation_2023}
Ferrero, S., Ceccarelli, C., Ugliengo, P., Sodupe, M., \& Rimola, A. 2023, The Astrophysical Journal, 951, 150

\bibitem[{Ferrero {et~al.}(2022)Ferrero, Grieco, Ibrahim Mohamed, Dulieu, Rimola, Ceccarelli, Nervi, Minissale, \& Ugliengo}]{ferrero_acetaldehyde_2022}
Ferrero, S., Grieco, F., Ibrahim Mohamed, A.-S., {et~al.} 2022, Monthly Notices of the Royal Astronomical Society, 516, 2586

\bibitem[{Fuchs {et~al.}(2009)Fuchs, Cuppen, Ioppolo, Romanzin, Bisschop, Andersson, Van~Dishoeck, \& Linnartz}]{Fuchs2009}
Fuchs, G.~W., Cuppen, H.~M., Ioppolo, S., {et~al.} 2009, Astronomy and Astrophysics, 505, 629

\bibitem[{Garcia‐Ratés \& Neese(2020)}]{garciarates_effect_2020}
Garcia‐Ratés, M. \& Neese, F. 2020, Journal of Computational Chemistry, 41, 922

\bibitem[{{Garrod}(2013)}]{Garrod2013}
{Garrod}, R.~T. 2013, Astrophys. J., 765, 60

\bibitem[{Garrod {et~al.}(2022)Garrod, Jin, Matis, Jones, Willis, \& Herbst}]{Garrod2022}
Garrod, R.~T., Jin, M., Matis, K.~A., {et~al.} 2022, The Astrophysical Journal Supplement Series, 259, 1

\bibitem[{Garrod {et~al.}(2007)Garrod, Wakelam, \& Herbst}]{garrod_non-thermal_2007}
Garrod, R.~T., Wakelam, V., \& Herbst, E. 2007, Astronomy \& Astrophysics, 467, 1103

\bibitem[{{Gerakines} {et~al.}(1995){Gerakines}, {Schutte}, {Greenberg}, \& {van Dishoeck}}]{Gerakines1995}
{Gerakines}, P.~A., {Schutte}, W.~A., {Greenberg}, J.~M., \& {van Dishoeck}, E.~F. 1995, \aap, 296, 810

\bibitem[{Gillan(1987)}]{Gillan1987}
Gillan, M.~J. 1987, Journal of Physics C: Solid State Physics, 20, 3621, publisher: IOP Publishing

\bibitem[{Guo {et~al.}(2018)Guo, Riplinger, Becker, Liakos, Minenkov, Cavallo, \& Neese}]{guo_communication_2018}
Guo, Y., Riplinger, C., Becker, U., {et~al.} 2018, The Journal of Chemical Physics, 148, 011101

\bibitem[{Herbst \& Van~Dishoeck(2009)}]{Herbst2009}
Herbst, E. \& Van~Dishoeck, E.~F. 2009, Annual Review of Astronomy and Astrophysics, 47, 427, publisher: Annual Reviews ISBN: 0066-4146

\bibitem[{{Hidaka} {et~al.}(2004){Hidaka}, {Watanabe}, {Shiraki}, {Nagaoka}, \& {Kouchi}}]{Hidaka2004}
{Hidaka}, H., {Watanabe}, N., {Shiraki}, T., {Nagaoka}, A., \& {Kouchi}, A. 2004, \apj, 614, 1124

\bibitem[{Holdship {et~al.}(2019)Holdship, Viti, Codella, Rawlings, Jimenez-Serra, Ayalew, Curtis, Habib, Lawrence, Warsame, \& Horn}]{Holdship2019}
Holdship, J., Viti, S., Codella, C., {et~al.} 2019, Astrophys. J., 880, 138

\bibitem[{{Hudson}(2017)}]{Hudson2017}
{Hudson}, R.~L. 2017, Spectrochimica Acta Part A: Molecular Spectroscopy, 187, 82

\bibitem[{Ibrahim {et~al.}(2022)Ibrahim, Guillemin, Chaquin, Markovits, \& Krim}]{ibrahim_formation_2022}
Ibrahim, M., Guillemin, J.-C., Chaquin, P., Markovits, A., \& Krim, L. 2022, Physical Chemistry Chemical Physics, 24, 23245

\bibitem[{Ibrahim {et~al.}(2024)Ibrahim, Guillemin, Chaquin, Markovits, \& Krim}]{ibrahim_significant_2024}
Ibrahim, M., Guillemin, J.-C., Chaquin, P., Markovits, A., \& Krim, L. 2024, Physical Chemistry Chemical Physics, 26, 4200

\bibitem[{Imai {et~al.}(2016)Imai, Sakai, Oya, López-Sepulcre, Watanabe, Ceccarelli, Lefloch, Caux, Vastel, Kahane, Sakai, Hirota, Aikawa, \& Yamamoto}]{Imai2016}
Imai, M., Sakai, N., Oya, Y., {et~al.} 2016, Astrophys. J., 830, L37

\bibitem[{{Jim{\'e}nez-Serra} {et~al.}(2016){Jim{\'e}nez-Serra}, {Vasyunin}, {Caselli}, {Marcelino}, {Billot}, {Viti}, {Testi}, {Vastel}, {Lefloch}, \& {Bachiller}}]{JImenezSerra2016}
{Jim{\'e}nez-Serra}, I., {Vasyunin}, A.~I., {Caselli}, P., {et~al.} 2016, \apjl, 830, L6

\bibitem[{Jim{\'{e}}nez-Serra {et~al.}(2021)Jim{\'{e}}nez-Serra, Vasyunin, Spezzano, Caselli, Cosentino, \& Viti}]{Jimenez-Serra2021}
Jim{\'{e}}nez-Serra, I., Vasyunin, A.~I., Spezzano, S., {et~al.} 2021, The Astrophysical Journal, 917, 44

\bibitem[{Jiménez-Serra {et~al.}(2016)Jiménez-Serra, Vasyunin, Caselli, Marcelino, Billot, Viti, Testi, Vastel, Lefloch, \& Bachiller}]{Jimenez_Serra_2016}
Jiménez-Serra, I., Vasyunin, A.~I., Caselli, P., {et~al.} 2016, The Astrophysical Journal, 830, L6, arXiv: 1609.05045 Publisher: American Astronomical Society

\bibitem[{Jin \& Garrod(2020)}]{Jin2020}
Jin, M. \& Garrod, R.~T. 2020, The Astrophysical Journal Supplement Series, 249, 26, arXiv: 2006.11127 Publisher: American Astronomical Society

\bibitem[{Jørgensen {et~al.}(2018)Jørgensen, Müller, Calcutt, Coutens, Drozdovskaya, Öberg, Persson, Taquet, Van~Dishoeck, \& Wampfler}]{jorgensen_alma-pils_2018}
Jørgensen, J.~K., Müller, H. S.~P., Calcutt, H., {et~al.} 2018, Astronomy \& Astrophysics, 620, A170

\bibitem[{K{\"a}stner(2014)}]{Kastner2014}
K{\"a}stner, J. 2014, Wiley Interdisciplinary Reviews: Computational Molecular Science, 4, 158

\bibitem[{K{\"a}stner {et~al.}(2009)K{\"a}stner, Carr, Keal, Thiel, Wander, \& Sherwood}]{kae09a}
K{\"a}stner, J., Carr, J.~M., Keal, T.~W., {et~al.} 2009, J. Phys. Chem. A, 113, 11856

\bibitem[{Knowles {et~al.}(1993)Knowles, Hampel, \& Werner}]{knowles_coupled_1993}
Knowles, P.~J., Hampel, C., \& Werner, H.-J. 1993, The Journal of Chemical Physics, 99, 5219

\bibitem[{Kozuch \& Martin(2011)}]{Kozuch2011}
Kozuch, S. \& Martin, J.~M. 2011, Physical Chemistry Chemical Physics, 13, 20104, publisher: The Royal Society of Chemistry

\bibitem[{Lamberts(2018)}]{lamberts_interstellar_2018}
Lamberts, T. 2018, Astronomy and Astrophysics, 615, L2

\bibitem[{Lamberts {et~al.}(2022)Lamberts, Fedoseev, Van~Hemert, Qasim, Chuang, Santos, \& Linnartz}]{lamberts_methane_2022}
Lamberts, T., Fedoseev, G., Van~Hemert, M.~C., {et~al.} 2022, The Astrophysical Journal, 928, 48

\bibitem[{Lamberts \& Kästner(2017{\natexlab{a}})}]{Lamberts2017a}
Lamberts, T. \& Kästner, J. 2017{\natexlab{a}}, The Astrophysical Journal, 846, 43, arXiv: 1708.05555 Publisher: IOP Publishing

\bibitem[{Lamberts \& Kästner(2017{\natexlab{b}})}]{Lamberts2017b}
Lamberts, T. \& Kästner, J. 2017{\natexlab{b}}, Journal of Physical Chemistry A, 121, 9736, publisher: American Chemical Society

\bibitem[{Lamberts {et~al.}(2019)Lamberts, Markmeyer, Kolb, \& K{\"a}stner}]{Lamberts2019}
Lamberts, T., Markmeyer, M.~N., Kolb, F.~J., \& K{\"a}stner, J. 2019, ACS Earth and Space Chemistry, 3, 958

\bibitem[{Maity {et~al.}(2015)Maity, Kaiser, \& Jones}]{maity_formation_2015}
Maity, S., Kaiser, R.~I., \& Jones, B.~M. 2015, Physical Chemistry Chemical Physics, 17, 3081

\bibitem[{Megías {et~al.}(2022)Megías, Jiménez-Serra, Martín-Pintado, Vasyunin, Spezzano, Caselli, Cosentino, \& Viti}]{megias_complex_2022}
Megías, A., Jiménez-Serra, I., Martín-Pintado, J., {et~al.} 2022, Monthly Notices of the Royal Astronomical Society, 519, 1601

\bibitem[{Meisner {et~al.}(2017)Meisner, Lamberts, \& Kästner}]{Meisner2017}
Meisner, J., Lamberts, T., \& Kästner, J. 2017, ACS Earth and Space Chemistry, 1, 399, arXiv: 1708.05559 Publisher: American Chemical Society

\bibitem[{Metz {et~al.}(2014)Metz, Kästner, Sokol, Keal, \& Sherwood}]{Metz2014}
Metz, S., Kästner, J., Sokol, A.~A., Keal, T.~W., \& Sherwood, P. 2014, Wiley Interdisciplinary Reviews: Computational Molecular Science, 4, 101

\bibitem[{Miksch {et~al.}(2021)Miksch, Riffelt, Oliveira, Kästner, \& Molpeceres}]{Miksch2021}
Miksch, A.~M., Riffelt, A., Oliveira, R., Kästner, J., \& Molpeceres, G. 2021, Monthly Notices of the Royal Astronomical Society, 505, 3157

\bibitem[{Molpeceres {et~al.}(2024{\natexlab{a}})Molpeceres, Furuya, \& Aikawa}]{Molpeceres2024com}
Molpeceres, G., Furuya, K., \& Aikawa, Y. 2024{\natexlab{a}}, Astronomy \& Astrophysics, 688, A150

\bibitem[{Molpeceres {et~al.}(2022{\natexlab{a}})Molpeceres, Jimenez-Serra, Oba, Nguyen, Watanabe, Garcia de~la Concepcion, Mate, Oliveira, \& Kästner}]{Molpeceres2022}
Molpeceres, G., Jimenez-Serra, I., Oba, Y., {et~al.} 2022{\natexlab{a}}, Astronomy \& Astrophysics, 663

\bibitem[{Molpeceres \& Kästner(2021)}]{Molpeceres2021}
Molpeceres, G. \& Kästner, J. 2021, The Astrophysical Journal, 910, 55

\bibitem[{Molpeceres {et~al.}(2022{\natexlab{b}})Molpeceres, Kästner, Herrero, Peláez, \& Maté}]{molpeceres_desorption_2022}
Molpeceres, G., Kästner, J., Herrero, V.~J., Peláez, R.~J., \& Maté, B. 2022{\natexlab{b}}, Astronomy \& Astrophysics, 664, A169

\bibitem[{Molpeceres \& Rivilla(2022)}]{molpeceres_radical_2022}
Molpeceres, G. \& Rivilla, V.~M. 2022, Astronomy \& Astrophysics, 665, A27

\bibitem[{Molpeceres {et~al.}(2023)Molpeceres, Rivilla, Furuya, Kästner, Maté, \& Aikawa}]{molpeceres_processing_2023}
Molpeceres, G., Rivilla, V.~M., Furuya, K., {et~al.} 2023, Monthly Notices of the Royal Astronomical Society, 521, 6061

\bibitem[{Molpeceres {et~al.}(2024{\natexlab{b}})Molpeceres, Tsuge, Furuya, Watanabe, San~Andrés, Rivilla, Colzi, \& Aikawa}]{molpeceres_carbon_2024}
Molpeceres, G., Tsuge, M., Furuya, K., {et~al.} 2024{\natexlab{b}}, The Journal of Physical Chemistry A, acs.jpca.3c08286

\bibitem[{Mondal {et~al.}(2021)Mondal, Gorai, Sil, Ghosh, Etim, Chakrabarti, Shimonishi, Nakatani, Furuya, Tan, \& Das}]{mondal_is_2021}
Mondal, S.~K., Gorai, P., Sil, M., {et~al.} 2021, The Astrophysical Journal, 922, 194

\bibitem[{Nagaoka {et~al.}(2007)Nagaoka, Watanabe, \& Kouchi}]{Nagaoka2007}
Nagaoka, A., Watanabe, N., \& Kouchi, A. 2007, The Journal of Physical Chemistry A, 111, 3016

\bibitem[{Neese {et~al.}(2020)Neese, Wennmohs, Becker, \& Riplinger}]{Neese2020}
Neese, F., Wennmohs, F., Becker, U., \& Riplinger, C. 2020, Journal of Chemical Physics, 152, 224108

\bibitem[{Nguyen {et~al.}(2019)Nguyen, Fourré, Favre, Barois, Congiu, Baouche, Guillemin, Ellinger, \& Dulieu}]{nguyen_formation_2019}
Nguyen, T., Fourré, I., Favre, C., {et~al.} 2019, Astronomy \& Astrophysics, 628, A15

\bibitem[{{Nguyen} {et~al.}(2023){Nguyen}, {Oba}, {Sameera}, {Furuya}, {Kouchi}, \& {Watanabe}}]{Nguyen2023}
{Nguyen}, T., {Oba}, Y., {Sameera}, W.~M.~C., {et~al.} 2023, \apj, 944, 219

\bibitem[{{Nguyen} {et~al.}(2021{\natexlab{a}}){Nguyen}, {Oba}, {Sameera}, {Kouchi}, \& {Watanabe}}]{Nguyen2021b}
{Nguyen}, T., {Oba}, Y., {Sameera}, W.~M.~C., {Kouchi}, A., \& {Watanabe}, N. 2021{\natexlab{a}}, \apj, 918, 73

\bibitem[{{Nguyen} {et~al.}(2021{\natexlab{b}}){Nguyen}, {Oba}, {Sameera}, {Kouchi}, \& {Watanabe}}]{Nguyen2021}
{Nguyen}, T., {Oba}, Y., {Sameera}, W.~M.~C., {Kouchi}, A., \& {Watanabe}, N. 2021{\natexlab{b}}, \apj, 922, 146

\bibitem[{{Nguyen} {et~al.}(2020){Nguyen}, {Oba}, {Shimonishi}, {Kouchi}, \& {Watanabe}}]{Nguyen2020}
{Nguyen}, T., {Oba}, Y., {Shimonishi}, T., {Kouchi}, A., \& {Watanabe}, N. 2020, \apjl, 898, L52

\bibitem[{Oba {et~al.}(2014)Oba, Osaka, Watanabe, Chigai, \& Kouchi}]{Oba2014}
Oba, Y., Osaka, K., Watanabe, N., Chigai, T., \& Kouchi, A. 2014, FaDi, 168, 185

\bibitem[{Oba {et~al.}(2018)Oba, Tomaru, Lamberts, Kouchi, \& Watanabe}]{Oba2018}
Oba, Y., Tomaru, T., Lamberts, T., Kouchi, A., \& Watanabe, N. 2018, Nature Astronomy, 2, 228

\bibitem[{Occhiogrosso {et~al.}(2014)Occhiogrosso, Vasyunin, Herbst, Viti, Ward, Price, \& Brown}]{occhiogrosso_ethylene_2014}
Occhiogrosso, A., Vasyunin, A., Herbst, E., {et~al.} 2014, Astronomy \& Astrophysics, 564, A123

\bibitem[{Papajak {et~al.}(2011)Papajak, Zheng, Xu, Leverentz, \& Truhlar}]{papajak_perspectives_2011}
Papajak, E., Zheng, J., Xu, X., Leverentz, H.~R., \& Truhlar, D.~G. 2011, Journal of Chemical Theory and Computation, 7, 3027

\bibitem[{{Perrero} {et~al.}(2023){Perrero}, {Ugliengo}, {Ceccarelli}, \& {Rimola}}]{Perrero2023}
{Perrero}, J., {Ugliengo}, P., {Ceccarelli}, C., \& {Rimola}, A. 2023, \mnras, 525, 2654

\bibitem[{Purvis \& Bartlett(1982)}]{purvis_full_1982}
Purvis, G.~D. \& Bartlett, R.~J. 1982, The Journal of Chemical Physics, 76, 1910

\bibitem[{Qasim {et~al.}(2020)Qasim, Fedoseev, Chuang, He, Ioppolo, van Dishoeck, \& Linnartz}]{Qasim2020}
Qasim, D., Fedoseev, G., Chuang, K.~J., {et~al.} 2020, Nature Astronomy, 4, 781, arXiv: 2004.02506

\bibitem[{Rimola {et~al.}(2014)Rimola, Taquet, Ugliengo, Balucani, \& Ceccarelli}]{Rimola2014}
Rimola, A., Taquet, V., Ugliengo, P., Balucani, N., \& Ceccarelli, C. 2014, Astronomy and Astrophysics, 572

\bibitem[{Rommel {et~al.}(2011)Rommel, Goumans, \& K\"astner}]{Rommel2011-2}
Rommel, J.~B., Goumans, T.~P., \& K\"astner, J. 2011, J. Chem. Theory Comp., 7, 690

\bibitem[{Santra {et~al.}(2019)Santra, Sylvetsky, \& Martin}]{santra_minimally_2019}
Santra, G., Sylvetsky, N., \& Martin, J. M.~L. 2019, The Journal of Physical Chemistry A, 123, 5129

\bibitem[{Schreiner {et~al.}(2011)Schreiner, Reisenauer, Ley, Gerbig, Wu, \& Allen}]{schreiner_methylhydroxycarbene_2011}
Schreiner, P.~R., Reisenauer, H.~P., Ley, D., {et~al.} 2011, Science, 332, 1300

\bibitem[{Scibelli {et~al.}(2021)Scibelli, Shirley, Vasyunin, \& Launhardt}]{Scibelli2021}
Scibelli, S., Shirley, Y., Vasyunin, A., \& Launhardt, R. 2021, Mon. Not. Royal Astron. Soc., 504, 5754

\bibitem[{Senevirathne {et~al.}(2017)Senevirathne, Andersson, Dulieu, \& Nyman}]{SENEVIRATHNE201759}
Senevirathne, B., Andersson, S., Dulieu, F., \& Nyman, G. 2017, Molecular Astrophysics, 6, 59

\bibitem[{Shingledecker {et~al.}(2018)Shingledecker, Tennis, Gal, \& Herbst}]{shingledecker_cosmic-ray-driven_2018}
Shingledecker, C.~N., Tennis, J., Gal, R.~L., \& Herbst, E. 2018, The Astrophysical Journal, 861, 20

\bibitem[{Simons {et~al.}(2020)Simons, Lamberts, \& Cuppen}]{Simons2020}
Simons, M.~A., Lamberts, T., \& Cuppen, H.~M. 2020, Astronomy and Astrophysics, 634, A52, arXiv: 2001.04895

\bibitem[{Song \& Kästner(2017)}]{song_tunneling_2017}
Song, L. \& Kästner, J. 2017, The Astrophysical Journal, 850, 118

\bibitem[{Truong \& Stefanovich(1995)}]{truong_new_1995}
Truong, T.~N. \& Stefanovich, E.~V. 1995, Chemical Physics Letters, 240, 253

\bibitem[{Vazart {et~al.}(2020)Vazart, Ceccarelli, Balucani, Bianchi, \& Skouteris}]{vazart_gas-phase_2020}
Vazart, F., Ceccarelli, C., Balucani, N., Bianchi, E., \& Skouteris, D. 2020, Monthly Notices of the Royal Astronomical Society, 499, 5547

\bibitem[{Walsh {et~al.}(2014)Walsh, Millar, Nomura, Herbst, Weaver, Aikawa, Laas, \& Vasyunin}]{walsh_complex_2014}
Walsh, C., Millar, T.~J., Nomura, H., {et~al.} 2014, Astronomy \& Astrophysics, 563, A33

\bibitem[{{Watanabe} \& {Kouchi}(2002)}]{Watanabe2002}
{Watanabe}, N. \& {Kouchi}, A. 2002, \apjl, 571, L173

\bibitem[{Watanabe {et~al.}(2006)Watanabe, Nagaoka, Hidaka, Shiraki, Chigai, \& Kouchi}]{WATANABE2006}
Watanabe, N., Nagaoka, A., Hidaka, H., {et~al.} 2006, P$\&$SS, 54, 1107, simulations in Laboratory

\bibitem[{Woon \& Dunning(1994)}]{Woon1994}
Woon, D.~E. \& Dunning, T.~H. 1994, The Journal of Chemical Physics, 100, 2975, publisher: American Institute of Physics ISBN: 0021-9606

\end{thebibliography}

\begin{appendix}
   \section{Determination of reaction barriers using a small 2 \ce{H2O} cluster model. Validity of the implicit surface approximation.} \label{sec:appA}

\begin{figure}[h]
    \centering
    \vspace{1.5em}
    \includegraphics[width=\linewidth]{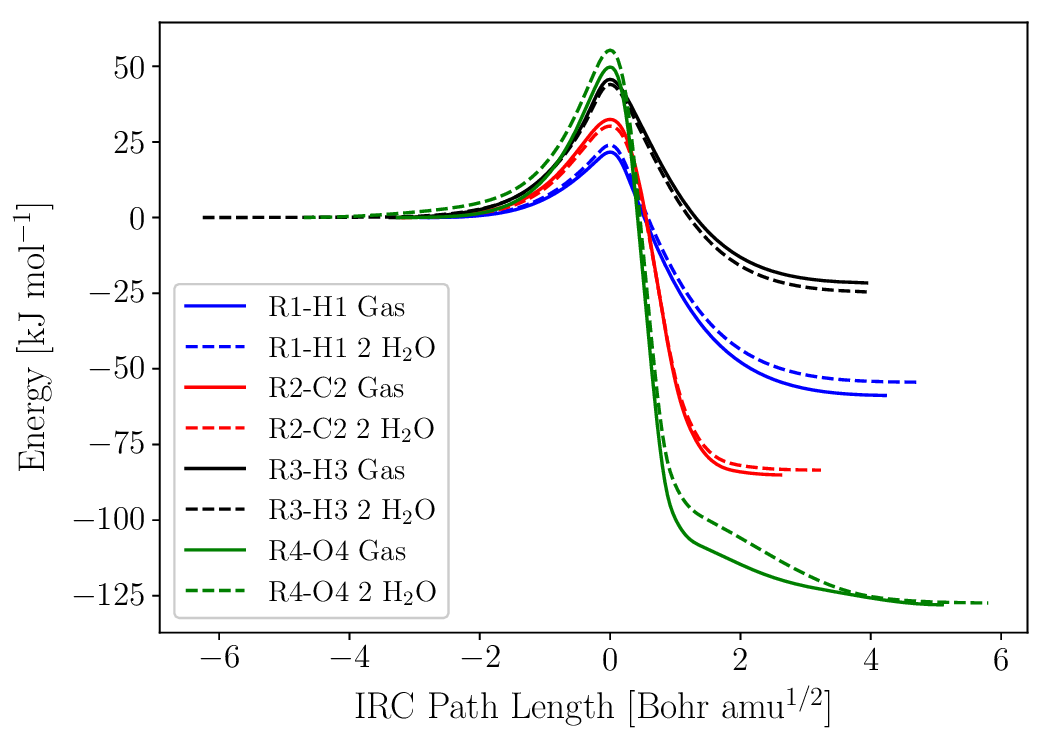} \\
    \caption{Equivalent to Figure \ref{fig:theo2}, but including reactions using an explicit 2 \ce{H2O} model (dashed line). }
    \label{fig:app1}
\end{figure} 

\begin{table}[bt!]
   \begin{center}
    \caption{Comparison in activation energies for the reactions \ref{eq:h1}-\ref{eq:c5} using a gas phase ($\Delta H^{ \ddagger}_{\textrm{Gas}}$ in kJ mol$^{-1}$) and an explicit two \ce{H2O} model ($\Delta H^{ \ddagger}_{\textrm{2 \ce{H2O}}}$ in kJ mol$^{-1}$).}
    \label{tab:energetics_appendix}
    \resizebox{\linewidth}{!}{
    \begin{tabular}{lccc}
   \toprule
    Reaction & Label & $\Delta H^{ \ddagger}_{\textrm{Gas}}$ &
    $\Delta H^{ \ddagger}_{\textrm{2 \ce{H2O}}}$ \\
   \bottomrule
   \ce{CH3CHO + H -> CH3CO + H2}   & \ref{eq:h1} &  18.5 & 19.6  \\
   \ce{CH3CHO + H -> CH3CH2O}      & \ref{eq:c2} &  31.0 & 27.7  \\
   \ce{CH3CHO + H -> CH2CHO + H2}  & \ref{eq:h3} &  41.2 & 37.1  \\
   \ce{CH3CHO + H -> CH3CHOH}      & \ref{eq:o4} &  47.3 & 48.1  \\
   \ce{CH3CHO + H -> CH4 + HCO}    & \ref{eq:c5} & 142.6 & 140.1  \\
    \bottomrule
   \end{tabular}
    }
\end{center}
\end{table}

In the main text, we obtain the reaction descriptors and subsequent kinetic rate constants using two implicit approximations. In addition to an implicit solvation approach, we assume that the ASW surface does not play a major role in the reaction, introducing surface effects \emph{via} rotational partition function fixing. These approximations are required for including nuclear quantum effects in the calculation of our rate constant while keeping an accurate potential (CCSD(T)/aug-cc-pVTZ//rev-DSD-PBEP86(D4)/jun-cc-pv(T+d)Z). The applicability of these approximations is ensured by a single condition, that the reaction descriptors are not affected by the water matrix. Such a condition is fulfilled normally in adsorbates bound to ASW \emph{via} weak physisorption forces, although it has been applied to molecules as tightly bound as OH \citep{Meisner2017, Lamberts2017a, Lamberts2017b, Molpeceres2021,Molpeceres2022,molpeceres_radical_2022,ferrero_formation_2023}. Acetaldehyde has one of the lowest binding energies found for any COM \citep{ferrero_acetaldehyde_2022,molpeceres_desorption_2022} so we do not expect a different behavior to the cases cited above. Nevertheless, we explicitly tested whether the introduction of explicit water affects the reaction descriptors and reaction profiles in our studied reactions. The tests we performed include comparing the activation energies, with and without water molecules, $\Delta H^{ \ddagger}_{\textrm{Gas}}$, and $\Delta H^{ \ddagger}_{\textrm{2 \ce{H2O}}}.$ \rev{For the 2 \ce{H2O} coupled cluster calculations, we used the domain local pair natural orbital (DLPNO) \citep{guo_communication_2018} version of CCSD(T), using a \texttt{TightPNO} localization scheme.}

The results of the tests are shown in Figure \ref{fig:app1} and Table \ref{tab:energetics_appendix}. Starting with the IRC profiles shown in Figure \ref{fig:app1} it is evident that in all cases the barrier shapes are almost identical, with subtle differences. These are, for example, that in the 2 \ce{H2O} model the IRC path extends beyond the gas phase one, a consequence of the interaction of the reacting molecules with the 2 \ce{H2O}. The most important deviation appears for reaction \ref{eq:o4} that, in the 2 \ce{H2O} model, experiences a restructuring of the water dimer to better accommodate the \ce{CH2CHO} radical. Nevertheless, the change in the profile is still very small. Similarly, the deviations in $\Delta H^{ \ddagger}$ are minimal and cannot be ruled out if the deviations are a consequence of the 2 \ce{H2O} model or by the DLPNO scheme used in the computation of the molecular energies. Overall, we find that including explicit water molecules does not affect our results. While the inclusion of further water molecules can also have an impact, this impact will not be larger than the one exerted by the closest water molecules. We remember that polar solvent effects are introduced in the main text utilizing an implicit solvation approach. Overall, we confidently conclude that the approximations in this work are justified in light of these results. 

\section{Modeling recomendations} \label{sec:recommendations}

The picture left by the combination of our calculations and experiments is a complex one, despite the relative simplicity of the \ce{CH3CHO} molecule. We think this is a trait common to all COMs (with more than one carbon atom) hydrogenations. As chemical complexity increases, the possibility of reaction branches starts to emerge, and hydrogenation experiments are harder to reconcile with theory because the former are suited to detect final products whereas the latter are more suited to investigate elementary processes. We feel that approaches like ours, merging experiments and theory are fundamental to securely identify representative interstellar reaction routes. 

However, the here studied reactions (or reaction schemes) seek to provide reliable data to feed astrochemical models, and in the particular case of this study, COMs and deuterium fractionation reaction networks, aiming to a better description of interstellar environments. Once the complexity of a reaction scheme starts to increase, as the one shown in this work, several assumptions and recommendations for astrochemical models need to be given. In this section we provide our guidelines for the treatment of \ce{CH3CHO} hydrogenation, justifying our decisions.\footnote{In addition to our considerations, the numerical values for the rate constants are gathered in \url{https://zenodo.org/records/14278652}} In the first place, and although the most exact approach is to include the exact rate constants for \ref{eq:c2}-\ref{eq:o4} and an accurate modeling of H atom diffusion, we consider safe deactivating the reactions in reaction networks. We base our recommendation on the limited conversion that we observe for \ce{CH3CHO}, which remains mostly unaltered in our experiments, an observation supported by the \ce{CH3CHO <=> CH3CO + H} cycle that we derive theoretically. The minor presence of \ce{CH3CHDOD} in our experiments, the only product that unambiguously can come from reaction \ref{eq:c2d} (as a deuterated proxy of reaction \ref{eq:c2}) also encourages us to consider the branching ratio of reactions other than reaction \ref{eq:h1} very low, and in the absence of further evidence, negligible. The second interpretation of our results concerns the branching ratios ($\alpha$ values) of reactions \ref{eq:ac-h3}-\ref{eq:ac-c5}. Rationalizing this choice is easy. If we consider that reaction \ref{eq:ac-c2} is the only reaction from the \ce{CH3CHO + H} then, the 90\% of remaining \ce{CH3CHO} in our hydrogenation experiment (Section \ref{sec:experiments:H}) must come from reaction \ref{eq:ac-c2} alone. The remaining 10\% should be distributed along the different reaction channels \ref{eq:ac-h3}-\ref{eq:ac-c5}. It is extremely difficult to determine $\alpha$ for these reactions experimentally or theoretically. We consider that the best compromise is to equally weigh the remaining 10\% along the three different reaction channels. Therefore we recommend setting $\alpha$=0.90 for \ref{eq:ac-c2} and 0.03 for \ref{eq:ac-h3}-\ref{eq:ac-c5}.

We would like to indicate that a certain degree of chemical intuition and data interpretation is needed to provide these modeling recommendations. Because our calculations show a significant resistance of \ce{CH3CHO} to hydrogenation and deuteration, we interpret our data based on this observation and the evidence from the quantum chemical calculations. Should minor channels along the hydrogenation sequence, for example, \ce{H2CCO} hydrogenation reforming \ce{CH3CHO} as reported in the literature \citep{ferrero_formation_2023,Fedoseev2022} have a larger contribution, several of the $\alpha$ values mentioned above can vary. Nevertheless, we expect these changes to be small.

\end{appendix}

\end{document}